\documentclass[sigconf,nonacm]{acmart}
\usepackage{booktabs} % For formal tables
\usepackage{booktabs,tabularx,xcolor}
\usepackage{array} % (optional, for new column types)
\usepackage[most]{tcolorbox}

\usepackage{exii-macros}

\usepackage[table]{xcolor}
\usepackage{booktabs}
\usepackage{tabularx}
\usepackage{array}
\usepackage{xcolor} % make sure this is in your preamble

\definecolor{indigo}{RGB}{75,0,130}
\definecolor{teal}{RGB}{0,128,128}
\definecolor{maroon}{RGB}{128,0,0}

\makeatletter
\g@addto@macro\normalsize{%
  \setlength\abovedisplayshortskip{-9pt}
  \setlength\belowdisplayshortskip{3pt}
}
\makeatother

\begin{document}

\tolerance=400

\title[Thinkink]{Thinkink: 2D Spatial Ink-native Interaction with LLMs}

\author{Mohammad Hasan Payandeh}
\orcid{0000-0001-7712-3701} 
\affiliation{%
  \institution{Cheriton School of Computer Science, University of Waterloo}
  \city{Waterloo}
  \country{Canada}
}
\email{mpayandeh@uwaterloo.ca}

\author{Daniel Vogel}
\orcid{0000-0001-7620-0541}
\affiliation{%
  \institution{Cheriton School of Computer Science, University of Waterloo}
  \city{Waterloo}
  \country{Canada}
}
\email{dvogel@uwaterloo.ca}

\author{Jian Zhao}
\orcid{0000-0001-5008-4319} 
\affiliation{%
  \institution{Cheriton School of Computer Science, University of Waterloo}
  \city{Waterloo}
  \country{Canada}
}
\email{jianzhao@uwaterloo.ca}


\renewcommand{\shortauthors}{Payandeh et al.}

\begin{abstract}

People often use handwritten notes and sketches to externalize ideas for ideation. To integrate large language models (LLMs) into this practice, we propose Thinkink. Prompts can be handwritten text or drawn sketches with LLM-generated responses visualized as ink-like text and sketches spatially integrated into a shared canvas. A semantic tree streamlines ink interpretation, and a lightweight UI provides explicit control using a state machine. The tool was designed using a three-stage process. A formative study (N=12) examined current practices with conventional and digital inking methods. The results informed a technical probe for a diagnostic study (N=6) identifying usability and human-LLM interaction challenges. This motivated the design of Thinkink, with a final study (N=10) examining how people incorporate it into their ideation practices. We contribute design implications and a tool for ink-native LLM interaction where the user and LLM write and draw in a shared 2D canvas.

\end{abstract}

\begin{teaserfigure}
  \centering
  \includegraphics[width=\linewidth]{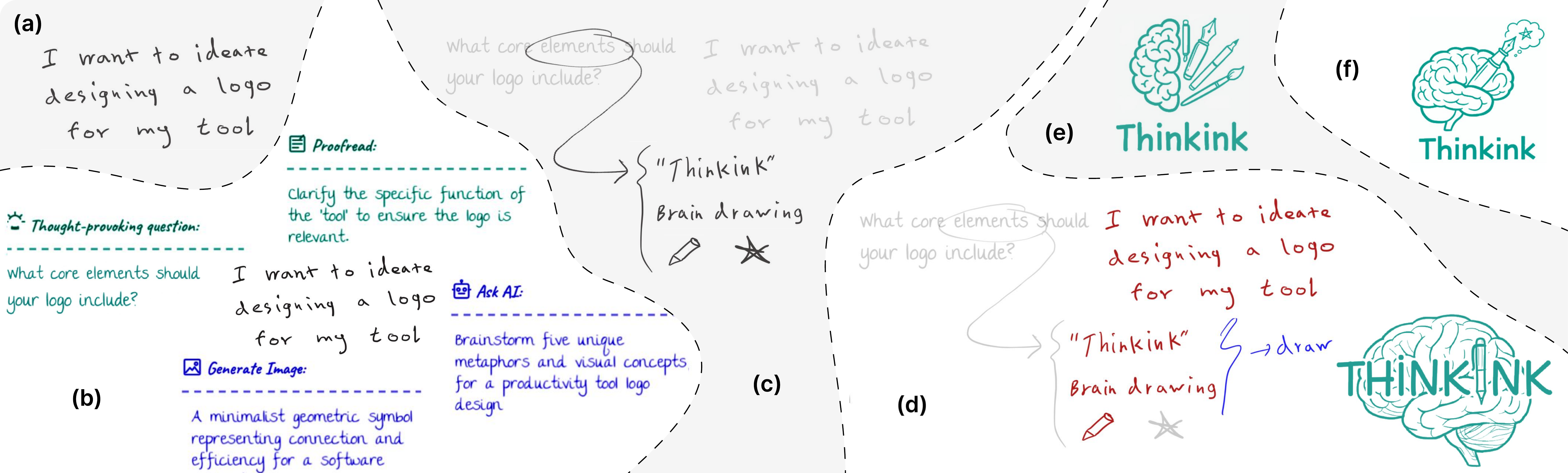}
  \caption{Example Thinkink usage scenario for logo ideation: (a) user writes high-level ideation goal; (b) system in `AI insights' mode responds with writing around the goal, including a reflective question, a proofreading point, a text generation request, and image generation request; (c) user considers reflective question and adds logo design ideas including `Thinkink', sketches of a pen and star, and `brain drawing' concept; (d) user switches to `Ask AI' mode, sketches an arrow, and writes `Draw', the system sets the context shown in red and responds with an initial logo concept; (e) user regenerates an alternative logo with the same context; (f) user adds the star to the context and regenerates a refined a logo concept.}
  \label{fig:teaser}
\end{teaserfigure}

\maketitle

\section{Introduction}

People often use handwritten notes and drawings to externalize their thoughts spatially on paper and digital devices~\cite{pittphilsci10414,Hutchins1995,Kirsh1999}.
This can support \emph{ideation} by helping people reflect, validate, and explore new thoughts~\cite{ExplanatoryReasoning1996,Spaceforthinking_jiansent2010}.
However, people can reach a deadlock and need support to explore new information, directions, and connections to move forward~\cite{Thagard1997Abductive,walton2014abductive,ideationrelyonabductive}.
LLMs can fill this gap by expanding the depth and breadth of existing thoughts~\cite{Tools_for_Thought2025, gu2023llmspotentialbrainstormingpartners}.
However, the typical chatbot-style of interaction does not match the non-linear two-dimensional nature of writing and drawing.

Prior work has explored LLM support for ideation~\cite{CHANG2025101755,AI-Augmented_Brainwriting2024,Ideation_with_Generative_AI2025}, sketch-based code editing~\cite{Code_Shaping2025}, ink-based note-taking~\cite{SkipWriter2024,Inkeraction2024,InkFM2025}, and sketching in general~\cite{SketchGPT2025,ImaginationVellum2025,SketchAgent2025}.
These systems demonstrate the value of LLM assistance for either inking or ideation, but they do not explore the intersection of the two.

Our research builds on these past work but is distinguished by its focus on using LLMs for ideation with all interaction entirely within an ink-native canvas.

To explore this space, we conducted a user-centered, iterative process including three stages.
Each stage investigated different questions, but they were connected through four cross-cutting aspects, consisting of \emph{Ideation workflow}, \emph{UI and Interaction Design}, \emph{LLM Integration}, and \emph{LLM Outputs}. 
Throughout this process, we developed a technical probe and then refined it into \textit{Thinkink}, a digital ink-native LLM-assisted ideation tool that enables people to interact with an LLM through ink input and receive ink-like responses directly on the canvas.

First, we conducted a formative focus group with 12 participants to understand \textit{RQ1: What user practices, needs, and expectations should inform the design of a digital ink-native LLM-assisted tool for ideation?}
Participants completed individual and collaborative ideation activities using their preferred media and reflected on their workflows. 
Based on participant feedback and observations, we derived four design guidelines.

In Stage 2, we translated these guidelines into a technical probe: a pen-first canvas with a semantic tree as backbone that interprets ink as drawings, concepts, requests, and generations, enabling context-aware, on-canvas LLM responses.  
Using the probe, we conducted a diagnostic study with 6 participants to understand \textit{RQ2: What human-LLM interaction challenges emerge when users engage with the technical probe?}
By observing participants using the probe across two ideation tasks and conducting follow-up interviews, we identified four key interaction design challenges.

In Stage 3, we addressed these challenges in refined prototype tool called Thinkink. It uses a state-machine interaction design that separates note-taking from LLM assistance, by making explicit prompting, insights, request iteration, and generation inspection distinct modes.  
We conducted a usage study with 10 participants to examine 
\textit{RQ3: How do users utilize Thinkink in practice?}
The results revealed four usage patterns that validate the Thinkink design and also suggest approaches and priorities for ink-native LLM interaction in general.

In summary, we make the following contributions:
\begin{itemize}
    \item Empirical knowledge that includes design guidelines, improvement areas, usage patterns, and discussions for building ink-native tools that unify inking workflows with LLM support, grounded in findings from a user-centered, iterative process. 
    \item \textit{Thinkink}, a digital inking tool for ideation that enables ink-based prompting and ink-like LLM responses directly on a 2D canvas. 
\end{itemize}

\section{Background and Related Work}

This section discusses related work on ideation, digital inking, and LLM-supported inking. 
We do not review the broader creativity-support literature spanning early uncertain phases to later well-defined stages, as our work focuses specifically on ideation.

\subsection{Ideation and Supporting Technologies}

Ideation is the process of generating new evidence or information that satisfies a goal, it is often described as divergent thinking or brainstorming~\cite{koestler1964act}. 
When people engage in ideation, they rely on \textit{abductive reasoning}, which means starting from evidence, usually incomplete, to find a plausible explanation~\cite{Thagard1997Abductive,walton2014abductive}. 
This is different than other types of reasoning, such as deductive reasoning which applies rules to derive necessary conclusions and inductive reasoning which generalizes from examples. 
In ideation, people specifically rely on abductive reasoning to externalize concepts, notice patterns, and identify new information, including new interpretations or next possibilities~\cite{ideationrelyonabductive}. 
Thus, ideation is not only idea generation, but also an abductive and explanatory process of making sense of partial evidence~\cite{ExplanatoryReasoning1996}.

Externalization enhances abductive reasoning~\cite{pittphilsci10414, Hutchins1995, Kirsh1999}. 
One approach to externalization is self-explanation, or ``Rubberducking'', where people explain their internal thoughts out loud to articulate assumptions and deepen understanding~\cite{Eliciting_self-explanations1994,RobotDuck2023}. 
Our focus is externalization using visuals by writing and sketching thoughts, an approach that works well for visual thinkers~\cite{Thagard1997Abductive, Spaceforthinking_jiansent2010}.
Both approaches help inspect and reorganize own thinking, but by definition they largely work with knowledge a person already has; they do not contribute new perspectives, analogies, or candidate directions.

Recent work uses LLMs to address this issue, framed as part of a broader agenda to create ``tools for thought'' that augment rather than replace cognition~\cite{Tools_for_Thought2025}. 
In ideation, LLMs can support divergence in three ways: they can broaden \emph{semantic breadth} by surfacing concepts from a wider conceptual space, suggest alternative framings or directions, and \emph{sustain divergence} by helping users exploring before converging too early on a single solution.

Prior work has been explored LLMs as brainstorming through prompt frameworks for eliciting creative thoughts~\cite{CHANG2025101755}, collaborative brainwriting 
, a group ideation method in which participants iteratively write and build on one another's ideas, as well as
idea evaluation~\cite{AI-Augmented_Brainwriting2024}, and role-based accounts of human--AI ideation 
that describe different divisions of labor, such as AI as the primary idea generator versus AI as support for human-led ideation~\cite{Ideation_with_Generative_AI2025}. 
~\cite{gu2023llmspotentialbrainstormingpartners}. 
Case studies have also shown how LLMs can act as brainstorming partners in complex domains such as mathematics, scientific research, and other technical problem-solving tasks~\cite{gu2023llmspotentialbrainstormingpartners}.
Collectively, this work suggests that LLMs can expand the space of possibilities considered during ideation.
However, interaction with LLMs remains largely conversational and text-centric, with ideas exchanged through prompts, chat, or documents. 
This mode of interaction can be suboptimal, especially for visual thinkers who externalize ideas through sketching and handwriting, as it requires constant back-and-forth between the inking canvas and the LLM interface. 
Our work fills this gap by intertwining people's sketches and handwriting with LLM-generated responses on a unified canvas.

\subsection{Digital Inking Tools}

Digital tools demonstrate how pen input (\ie{} the ``ink'') can enhance note-taking, sketching, and thinking through direct, low-friction interaction. 
They have shown that ink is not only a recording medium, but also an interaction primitive. 

For example, ActiveInk~\cite{ActiveInk2019} supports sensemaking by letting users annotate data and then activate those strokes for analytic actions, linking externalization and interaction. 
Work on Sketchnoting~\cite{Sketchnote2021} further characterizes how people combine text, diagrams, layout, and styling in visual notes, highlighting the importance of structure, recomposition, and lightweight expressiveness in digital tools. 
These systems establish key design principles for pen-based environments: preserve the immediacy of ink, support mixed visual representations, and keep interaction close to the workspace.

InkSeine~\cite{InkSeine2007} uses existing ink to initiate in-situ search, treats queries as first-class objects, and lets users fluidly interleave note-taking with retrieval~\cite{case2016looking}. 
It shows how ink can act as a simple input command without forcing users to leave the canvas. However, InkSeine routes those interpretations into a separate search interface rather than using the canvas itself as a shared space for AI-mediated ideation and response.
We are inspired from this work by using ink as input, but we also integrate the resulting information in an ink-like form directly within a 2D spatial canvas, not in a separate interface. 
This demands careful interaction design, as a single workspace must support both traditional inking actions, such as writing, erasing, and moving drawings, and AI-specific actions, such as writing prompts, erasing prompts, and generating responses, while remaining easy to understand and fluid to use during open-ended exploration~\cite{Pen_+_touch2010,WritLarge2017}.

More broadly, whereas prior digital inking tools treat ink primarily as user-authored input for note-taking, search, or analytic commands, we treat the canvas as a shared ideation space in which AI also contributes directly in ink.

\subsection{AI-powered Digital Inking}

A growing body of prior work utilize AI/LLMs combined with digital ink, but for tasks other than ideation. 
Some systems improve writing itself: SkipWriter~\cite{SkipWriter2024} uses LLMs for abbreviated handwritten text input, or \citet{Handwriting_Enhancement2024} propose handwriting enhancement techniques that beautify handwritten text while preserving personal style.
Others focus on recognizing and manipulating ink. 
Inkeraction segments and classifies handwritten objects, identifies relationships, and synthesizes strokes to support editing and writing workflows~\cite{Inkeraction2024}, while InkFM provides a foundation for full-page handwritten note understanding across text, math, and drawings~\cite{InkFM2025}. 
These works share an important technical foundation, namely the ability of AI to interpret ink correctly. 
However, when they model relationships, the semantics are typically relatively simple and static; they are less suited to interpreting the evolving, context-dependent meaning of mixed sketches and notes during open-ended ideation (\eg{} a circle with a line under it can be a balloon, but the same circle, when the user draws a hand for the line, is the head of a stickman).

A few systems come closer to our work. 
Code Shaping shows that free-form sketches can communicate editing intent to AI, but in the specialized context of code manipulation~\cite{Code_Shaping2025}.
Visual Sketchpad gives multimodal language models an internal visual chain-of-thought~\cite{Visual_Sketchpad2024} aimed at improving model reasoning, not supporting end-user ideation on a shared canvas. 
SketchGPT lets users interact with LLMs using sketches and speech to generate context-aware responses~\cite{SketchGPT2025},
producing confirmable operation lists on the underlying UI.
Common across all three works, the interaction is mainly one-way, from user ink to system interpretation, and responses are delivered through interface actions rather than as ink-native contributions within the same 2D workspace.
ImaginationVellum presents a generative-AI canvas where the inks in a 2D workspace acts as a prompt for visual generation~\cite{ImaginationVellum2025}, but its domain is image ideation rather than ink-based conceptual exploration. 
SketchAgent supports conversational sequential sketch generation and refinement~\cite{SketchAgent2025}, but it primarily completes or modifies a drawing rather than helping users explore divergent idea spaces.

What is missing is an ideation-focused LLM that shares the page: it incrementally grounds in the evolving, spatial semantics of mixed ink and responds in ink (placing lightweight annotations/sketches directly in the same 2D workspace), rather than only triggering UI actions or generating standalone images.

Our work brings together ideation with ink-native support. Unlike prior ideation tools, it is not primarily chat-based; 
unlike prior ink systems, it is not limited to recognition, beautification, or command invocation. Instead, it treats ink as both a cognitive artifact and an interaction medium for generative exploration, enabling users and LLM to co-develop ideas directly on the same canvas.

\section{Formative Study}

To address \emph{RQ1: What user practices, needs, and expectations should inform the design of a digital ink-native LLM-assisted tool for ideation?} we conducted a formative study with a focus group~\cite{focusedgroupstudy2006}.  
Our goal was to identify design guidelines for designing a technical probe as an initial prototype.

\subsection{Method}

We conducted a 2-hour in-person focus group with 12 participants from an HCI laboratory (\textit{M}=7, \textit{F}=3, \textit{ND}=2; age 23--40). 
The cohort included two professors, and ten Master's and PhD students. 
Four participants listed data visualization as their primary research area, one worked in XR/VR, one in interactive art and technology, and the remainder indicated general HCI. 
Participants took part voluntarily and did not receive any compensation.

The session combined an externalization activity with a group reflection. 
After a short introduction and consent process, participants read a brief sheet describing question categories and examples that encourage ideation (See Appendix~\ref{apx:Formative_Question_Categories}). 
Each participant then selected or authored a question to answer individually and, optionally, another for pair work (See Appendix~\ref{apx:Formative_Participants_Choices}). 
They were asked to write the question, list at least five thought or evidence they have in their mind, externalize these evidence visually, think, explore, and draft an answer. 
Participants could use any medium they preferred, including pen and paper, digital inking tools, or other non-inking digital tools such as Excalidraw~\cite{excalidraw}, draw.io~\cite{drawio}, or Figma~\cite{figma}, and they were also allowed to consult LLMs such as ChatGPT~\cite{chatgpt}. 
We supplied paper, pens, sticky notes, and colored pencils. 
The individual activity lasted approximately 20 minutes, followed by an optional 20-minute collaborative activity and a 20-minute semi-structured group discussion (See Appendix~\ref{apx:Formative_Discussion_Points} for the discussion points). 
Discussion points elicited contributions from multiple participants; quotes are reported without identifiers due to the shared recording setup.
We collected photographs or screenshots of the final artifacts, resulting in 15 sketches: 12 on paper, 2 created using a digital inking tool, and 1 in Figma; 12 individual and 3 collaborative
(see Appendix \ref{apx:Formative_Participants_Choices} for table of participant medium choices and \ref{apx:Formative_Artifacts} for the artifacts).

\subsection{Design Guidelines}

We transcribed the discussion verbatim and conducted an inductive thematic analysis~\cite{thematic} of the transcript together with an inspection of the artifacts. 
Four different high-level aspects served as analytic lenses throughout our process: \emph{Ideation workflow} concerns how users externalize, organize, and develop ideas; \emph{UI and Interaction Design} concerns the pen-first interactions and interface structures that support this process; \emph{LLM Integration} concerns how and when LLM capabilities are invoked within the digital inking tool; and \emph{LLM Outputs} concerns what the LLM generates and how those responses are presented on the canvas.

\subsubsection*{\textbf{Ideation workflow: Support Fluid Ideation Through Evolving Externalization Workflows}}
Participants described externalization as generative rather than merely representational: \pquote{I definitely think externalizing context is helpful … it helps me to open my mind}, and \pquote{I personally feel like once I start visualizing this causation, I start brainstorming more causations … expanding this}. 
However, their workflows were rarely linear. Some struggled to begin (\pquote{It’s hard for me to find a starting point … how would I visualize my thoughts?}), some moved from text to visuals (\pquote{You first write down the context, and then visualize}), and others felt a fixed sequence was constraining (\pquote{I follow the strict process … and I feel like that kind of restricts me}).
Several also found it difficult to decide on an appropriate representation (\pquote{I couldn’t really think about how to visualize it}). 
Together, these findings suggest that an ink-native LLM-assisted ideation tool should support evolving, non-linear externalization: users should be able to begin with partial text, rough marks, or simple structures, receive lightweight scaffolds when needed, and continuously revise and reorganize the canvas as ideas develop.

\subsubsection*{\textbf{UI and Interaction Design: Enable a Minimal, Paper-Like Inking Experience}}
Medium choice strongly favoured paper: 12 of the 15 artifacts were created on pen and paper. Participants associated paper with low distraction and immediacy, noting that \pquote{Pen and paper is less distracting … I lose my train of thought more easily with software} and \pquote{I could get many notifications on an iPad}. 
Digital tools were mainly valued for their malleability, as one participant noted: \pquote{iPad is quite limited … but moving things around is super easy}. 
Artifacts also relied mostly on simple diagrammatic elements such as arrows, circles, rectangles, clouds, and handwritten labels, with colours appearing in only 2 of the 15 artifacts (see Appendix~\ref{apx:Formative_Artifacts}). 
A digital ink-native LLM-assisted tool should therefore minimize interface chrome and preserve a paper-like experience: a full-screen canvas, pen-first interaction, simple strokes and basic shapes, and lightweight touch-based manipulation that retains digital flexibility without interrupting thought.

\subsubsection*{\textbf{LLM Integration: Integrate LLMs as an Ideation Partner Within the Canvas}}
Many participants said they view chatbots such as ChatGPT as tools for ideation and filling knowledge gaps, rather than as replacements for their own reasoning. Some used chatbots in the study.
As one participant explained, \pquote{the AI (LLMs) helped me to really get to think deeper and if I get stuck then I have help}. 
At the same time, participants noted that this support may be more helpful in domain-specific contexts: \pquote{if you’re doing something…highly specialized…you might get as much help}. 
This encourages integrating LLMs as a contextual ideation partner that works from the evolving sketch itself, helping users expand, connect, and elaborate ideas in place. 
Rather than functioning as a separate text-only assistant, the LLMs should be grounded in the current canvas state and contribute support that remains tightly coupled to the user’s ongoing externalization.

\subsubsection*{\textbf{LLM Outputs: Favour Exploratory, Question-Asking LLM Outputs Over Direct Answers}}
Participants were cautious about LLM Outputs that could bias their thinking or narrow the space of possibilities. 
Instead of direct answers, some of them preferred prompts that stimulate exploration, as reflected in the suggestion: \pquote{What if it[LLMs] can provoke us, ask us some questions…that will be more helpful in terms of opening up}. 
For digital ink-native ideation tools, LLM Outputs should therefore prioritize exploratory support, such as reflective questions, alternative perspectives, tensions, and missing considerations, over declarative answers. This style of output is better aligned with sketch-based ideation because it expands the search space, preserves user agency, and helps users continue reasoning through the canvas rather than deferring to the model.

\section{Technical Probe}

To explore how LLM support can be integrated directly into an inking workflow, we developed a technical probe using a Vite-based React and TypeScript canvas application~\cite{vite,react,typescript}. 
The probe maintains a semantic tree behind the scenes (unseen by users) as a backend structure. It leverages LLMs (Gemini 3 Flash Preview~\cite{gemini_api}) to continuously analyze user drawings and translate the evolving workspace into a structured representation, enabling the system to better understand user input and respond to requests.
This representation enables the system to interpret what the user has drawn and subsequently provide context-aware LLM responses directly on the canvas (Figure~\ref{fig:techincal_probe_semantictree2}).
Interaction traces are persisted using a lightweight Express logging server~\cite{express}.

Following the \emph{fluid externalization ideation workflow} guideline form the formative study, the technical probe  supports externalization by letting users begin with partial text, rough sketches, or simple marks and then continue adding, moving, and refining content on an open canvas without following a fixed sequence. After short idle periods, the system re-analyzes the workspace. 
Following the \emph{minimal paper-like inking UI} guideline, it preserves a minimal, paper-like interaction style through a full-screen drawing surface, direct pen input, and lightweight touch gestures for navigation and editing rather than menus or chat panels. 
Following the \emph{on-canvas LLM ideation partner} guideline, it integrates LLMs as an ideation partner within the canvas by grounding requests in specific drawings and returning responses as on-canvas textual annotations positioned near the relevant part of the sketch. 
Finally, following the \emph{exploratory, question-first LLM outputs} guideline, the probe emphasizes exploratory support by recognizing opportunities for follow-up questions, information retrieval, or suggestions for what to draw next, enabling the system to extend the user’s thinking in place rather than only delivering detached answers.

\subsection{User Interface}

The probe interface is a large, infinite white canvas with a subtle background grid and no visible UI buttons, so user drawings and LLM-generated content coexist directly on the workspace. 
User strokes remain primary in black, while LLM outputs are positioned near the triggering sketch so they read as spatial additions rather than separate chat messages; 
LLM generations they as tentative, semi-transparent annotations in two forms: teal for generation without request and indigo for requests. 
Tapping a request produces its response in teal. 
Tapping any generated content (with or without a request) finalizes it by turning it black and merging it into the evolving artifact, preserving the immediacy of paper while keeping the model’s contributions legible and integrated. 
Interaction is optimized for pen-and-touch on a tablet: the pen is used for drawing and erasing, two-finger gestures pan and zoom, and a one-finger double tap toggles eraser mode before erasing with the pen.

\subsection{Semantic Tree}

\begin{figure}
    \centering
    \includegraphics[width=\linewidth]{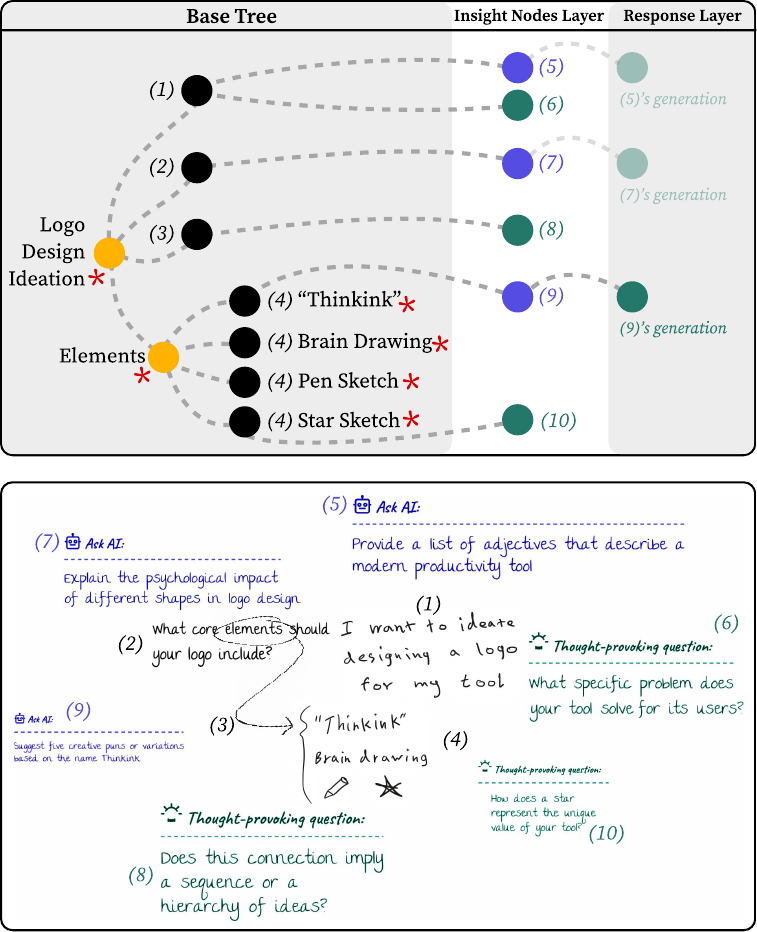}
    \caption{semantic tree of the user canvas; black=drawings, orange=concepts, indigo=requests, teal=generations with/without request; red stars mark the context selected to produce generation (9) from request (9).}
    \label{fig:techincal_probe_semantictree2}
    \vspace{-4mm}
\end{figure}

Behind the UI, the probe maintains a semantic tree that helps the system better interpret the user drawings, as shown in Figure~\ref{fig:techincal_probe_semantictree2}. 
The tree serves as an intermediate structure between raw ink and LLM prompting: rather than asking the model to reason over isolated strokes or a flat image alone, the system organizes the workspace into meaningful units and relations. 
Its \textit{Base Tree} is produced using the prompt in Appendix~\ref{apx:base-semantic-tree-generation}. 
Black nodes denote drawing nodes and orange nodes denote concept nodes. 
The tree’s \textit{Base Tree} is produced using the prompt in Appendix~\ref{apx:base-semantic-tree-generation}. Black nodes denote drawing nodes and orange nodes denote concept nodes. 

Drawing nodes represent individual pieces of ink, such as handwritten words, arrows, circles, or other sketch elements, while concept nodes represent higher-level groupings of related drawings or subclusters. 
Because users may work in several disconnected regions of the canvas, the representation allows multiple roots rather than forcing the entire workspace into a single hierarchy.

On top of the Base Tree, the system adds an \textit{Insight Nodes Layer} using the prompt in Appendix~\ref{apx:proactive-node-generation}. 
This layer introduces indigo request nodes, which mark places where the system identifies an opportunity to provide textual support, and teal generation-without-request nodes, which represent unsolicited model outputs (e.g., Socratic questions or suggestions for what to draw next).

Finally, the system adds a \textit{Response Layer}, which includes generations produced in response to request nodes (i.e., generation-with-request nodes attached to their corresponding requests). 
To produce generation-with-request nodes, the system utizlize user request and the context related to the request as a prompt for node generation (see Appendix~\ref{apx:text-response-generation}). 
The request is gathered by analysis of drawings. 
The context is gathered from the request's local and hierarchical neighborhood, such as its parent, siblings, relevant ancestors, and related sibling branches.
For example, Figure~\ref{fig:techincal_probe_semantictree2} shows dark red stars indicating the context selected for request node (9)---the drawing and concept nodes from the tree used to ground its generation. 
This layered structure clarifies the distinction between the user's drawings, the system's interpretation as requests for support, and the model's subsequent contributions to the canvas.

The tree is built and updated through an analysis pipeline that runs whenever the user becomes idle. 
The system first sends the model a composed view of the canvas together with individual drawings and asks it to generate the Base Tree (Appendix~\ref{apx:base-semantic-tree-generation}). It then prompts the model to augment that structure with insight nodes (Appendix~\ref{apx:proactive-node-generation}) and, when a request is present, to produce the corresponding Response Layer node using the prompt in Appendix~\ref{apx:text-response-generation}. The resulting text is placed near the relevant drawing, and if the user accepts it, it becomes part of the evolving workspace and is incorporated into subsequent analyses. In this way, the semantic tree allows the probe to treat ink not only as display content, but also as a structured input language for LLM insights.

\section{Diagnostic Study}

To address \emph{RQ2: What human-LLM interaction challenges emerge when users engage with the technical probe?} we conducted a diagnostic study with the technical probe. 
Our goal was to identify key interaction challenges before building Thinkink, exploring four aspects of the design.
Our findings revealed challenges in three aspects, while \emph{Ideation workflow} showed no issues, indicating that participants were satisfied with the technical probe's support in fluid ideation through evolving externalization workflows.

\subsection{Method}

We recruited 6 participants from a university community (\textit{M}=4, \textit{F}=2; age 26--37), each from a different domain: quantum information and computation, HCI, computational arts/game design, fluid mechanics, reinforcement learning and AI ethics, and vision science. 
The cohort included one postdoctoral researcher, three PhD students, and two Master's students. 
Participants received \$30 compensation. 
Overall, participants reported frequent use of pen and paper for thinking and ideation (\textit{M}=3.67, \textit{SD}=1.03 on a 5-point scale), infrequent use of digital inking tools (\textit{M}=2.50, \textit{SD}=0.84), and high use of LLMs for thinking and ideation (\textit{M}=4.67, \textit{SD}=0.52). 

The study was conducted in person on a 13-inch iPad Pro using our technical probe. 
Each session lasted approximately 90 minutes. 
After consent, demographics, and pre-task questions, participants completed a short tutorial. 
They then completed two ideation tasks: one based on a question from the same set used in the formative study (see Appendix~\ref{apx:Formative_Question_Categories}) and one related to their own area of expertise, chosen by the participants themselves. 
For each task, participants spent 10 minutes to externalize what they knew, explore and ideate with the LLM, and then reach the best conclusion they could. 
We collected think-aloud data, screen recordings, audio, interaction logs, and final artifacts  (see Appendix~\ref{apx:Diagnostic_Artifacts}), followed by a semi-structured interview and an 11-item 5-point Likert questionnaire. 
The full study materials are provided in Appendix~\ref{apx:Diagnostic}.

\subsection{Challenges}

We analyzed the qualitative data using the same method as in the formative study, employing the four high-level design aspects as analytical lenses. 
We further enriched our analysis with interaction logs, artifacts, questionnaire ratings, and semantic tree inspections.

Participants highlighted several strengths of the probe that echoed our formative design guidelines. Following the ``Ideation workflow'' guideline form the formative study, they valued the spatial canvas for maintaining a big-picture view of their evolving externalization: \pquote{with the canvas, I zoom out and see what happened}{P6}. 
Following the ``UI and Interaction Design'' guideline, they appreciated the minimal, no-UI design for staying out of the way of thinking: \pquote{very minimal ... I'm a fan of minimal stuff}{P5}. 
Following the ``LLM Integration'' guideline, participants found that even unconfirmed Insights could still spark further exploration: \pquote{The good thing about that is that when it offers, like, a suggestion, I don't need to take it as is....}{P2}. 
Following the ``LLM Outputs'' guideline, participants also saw value in Socratic questioning that expands the ideation space, suggesting that \pquote{[i like that] every time I ask a question, it responds with, like, another question....}{P2}.

The study also revealed challenges, reflected in participants’ interviews and survey responses (Figure~\ref{fig:survey_summary}; Appendix~\ref{apx:survey_questions_responses}). 
\subsubsection*{\textbf{UI and Interaction Design: Unclear Boundaries Between Note-Taking and LLM Modes, Causing Interaction Confusion}}
Although most participants appreciated the minimal interface, the lack of explicit cues also created confusion. 
Participants often could not tell when they were simply writing notes and when the system had started interpreting their writing as input for LLMs response: \pquote{Whenever I was writing something, I was thinking if the model is... considering... everything}{P1}. 
They were also sometimes unsure how to perform different actions, especially when interactions overlapped: \pquote{when I tried to move the screen... it moved some part of that text that I wrote before....}{P6}. 
These difficulties led participants to ask for more explicit interface support, including visible controls \pquote{I would honestly prefer having a toolbar....}{P2}, fewer and more clearly separated modes \pquote{having two different proactive and active sessions, that would make it easier....}{P4}, and lightweight onboarding embedded in the interface: \pquote{for any new application... they have kind of introduction... that would be more helpful...}{P6}.
\subsubsection*{\textbf{UI and Interaction Design: Limited Error Prevention, Recovery From Errors, and Control}}
Participants encountered a system that neither prevented common errors nor supported smooth recovery once errors occurred. 
Editing was brittle: \pquote{there's no razor, so it's a little bit painful that you cannot immediately erase it....}{P1}, \pquote{I feel it should have a control Z. A button...}{P4}, \pquote{Yeah, I need to have an option for undoing… the process....}{P6}, and \pquote{because of 2sec merging, the writing seperated to two text (which shouldn’t)...}{P1}.
Also, LLM Outputs were difficult to preserve or revisit, as indicated by \pquote{Maybe if that would have stayed a bit longer, so that I can choose to stay there....}{P4} and \pquote{when I press it, and then I continue writing, it disappears....}{P3}. 
Users also wanted more control over automation, noting premature intervention (\pquote{I was trying to actually write very fast... the AI (LLM) might start thinking about that part, and give me...}{P1} \pquote{I don't want the AI (LLM) to examine everything as soon as I'm done....}{P4}), and opaque and uneditable context selection for output generation (\pquote{It was not clear to me which part of the text my question, the AI (LLMs), is considering....}{P1} and \pquote{what if it caused… this mistake caused it to have the wrong prompt in future?...}{P4}).

\subsubsection*{\textbf{LLM Integration: Inability to Interact Directly with LLMs}}
Participants wanted an explicit way to prompt the LLMs in addition to receiving Insights, because autonomous feedback did not always match their immediate intent. 
Several emphasized the need to decide when the system should engage (\pquote{it is nice that it starts thinking when I am saying, hey, start thinking now....}{P1} and \pquote{I would prefer to write everything down... then... now I want you to examine this process...}{P4}), while others were frustrated when direct queries led to more questioning rather than answers (\pquote{I feel that it's, like, keep asking other questions... it's not giving any answer....}{P4}, and \pquote{I wouldn't prefer questions, honestly, in this stage....}{P3}). 

\subsubsection*{\textbf{LLM Outputs: Misalignment Between LLM Output Types/Formats and User Needs}}
LLM Outputs sometimes failed to match the representational needs of participants’ work. Users expected the system to respond visually when appropriate, e.g., \pquote{it would be really cool if it starts, like, sketching with me...}{P2}, \pquote{if it could draw some diagrams for me, some shapes for me. it would be top-notch...}{P5}, and \pquote{Do you want to draw a cube? I can draw a nice cu...}{P4}. 
This mismatch was especially evident in technical tasks, where plain-text responses were insufficient (\pquote{the presentation format of the AI (LLM) responses, if it is just text, is terrible}{P1}) and where unreadable formatting and rigid typography further reduced usefulness, as reflected in \pquote{It's not easy to read the latex format....}{P4} and \pquote{Can I adjust the size of the text, or no?}{P2}.

\subsection{Improvement Areas}

From these challenges, we derive the following improvement areas to inform the refinement of the technical probe.

\subsubsection*{\textbf{UI and Interaction Design: Make System Modes Explicit and Constrain Interactions Within Each Mode}}
The interface should make writing, editing, selection, and LLM-based modes clearly distinct and easy to switch between, so users can understand what the system is attending to and what actions are currently possible. Limiting each mode to a small, well-defined set of interactions, supported by lightweight cues such as visible controls, mode indicators, and brief onboardingm, would reduce accidental actions, clarify system behavior, and improve learnability without sacrificing interface simplicity.

\subsubsection*{\textbf{UI and Interaction Design: Prioritize Recoverability and Controllable Automation}}
Hybrid ink--LLM systems should be designed around reversibility, persistence, and inspectability. Robust undo/redo, editable merge outcomes, persistent response history, explicit context scoping, and manual LLM triggering would help users prevent errors, recover from them quickly, and override automation when it acts at the wrong time or with the wrong context.

\subsubsection*{\textbf{LLM Integration: Enable Both Explicit Prompting and Insights}}
The system should distinguish between on-demand prompting and Insights as two explicit but interoperable modes. Allowing users to manually invoke the LLM, while separately controlling Insights, would preserve user agency and reduce interruptions without sacrificing serendipitous assistance.

\subsubsection*{\textbf{LLM Outputs: Align Content and Types with Users Demands}}
LLM responses should adapt in both content and presentation. Beyond text, outputs can include proofreading and critique-oriented feedback, ink-like sketches, rendered mathematics, Markdown, and context-aware typography, ensuring they remain clear and useful across technical and presentation-oriented workflows.

\section{Thinkink}

\subsection{Overview}

\textit{Thinkink} is the refined version of our technical probe, redesigned in response to the diagnostic study (Figure~\ref{fig:teaser}; see Appendix~\ref{fig:screenshots-statemachine} for all screenshots).
The main changes are: adding stronger mechanisms for control and recovery such as confirmation, history, and manual context adjustment, separating explicit LLM prompting from Insights, expanding the types and formats of LLM Outputs, and making system modes explicit through a state machine interaction design~\cite{da2003improving, visualparadigm} (overview in Figure~\ref{fig:state-machine-overview}; detailed version in Appendix~\ref{fig:screenshots-statemachine}).
In contrast to the probe, Thinkink does not rely on loosely inferred LLMs interaction alone; instead, it structures interaction into clear substates that make available actions, transitions, and LLM behaviour more legible.

Thinkink retains the semantic tree as its backbone, but builds it on demand rather than regenerating the entire structure after every canvas change, as in the technical probe. 
Considering the state machine design shown in Figure~\ref{fig:state-machine-overview}, entering ``Prompt'' or ``Insights'' from ``Drawings'', ``Prompt'', or ``Insights'' regenerates only the \emph{base tree}. 
Entering ``Insights'' from ``Drawings'' or ``Prompt'' adds the \emph{Insight Nodes Layer} on top of the \emph{base tree}. 
Entering ``Iterate'' adds the \emph{Response Layer}, but only includes a generation with request node for the active request: if reached from ``Insights'', the request is the selected request node from the \emph{Insight Nodes Layer}; if reached from ``Prompt'', it is the explicit request drawing, attached as a request node to the relevant drawing.
Constructing the semantic tree incrementally reduces the number of LLM calls and the input tokens per call, thereby decreasing user wait times.

\subsection{User Interface}

Thinkink incorporates mechanisms for user control and recovery during interaction with LLMs. 
Users can compose explicit on-canvas requests, browse Insights, and manually adjust selected drawings by the model as context used for response generation. 
LLM Outputs are no longer treated as only temporary overlays: they can be inspected, browsed as alternatives, accepted into the workspace, and revisited through history. 
A history bar supports both revision and exploration by enabling users to move backward and forward through edits, revisit earlier LLM Outputs, and trigger regeneration of new alternative responses or Insights. During regeneration, previous outputs are included in the prompt so the model can produce different alternatives rather than repeating prior ones, which better supports ideation.

Thinkink retains the probe's pen-first infinite canvas and minimal visual style, but adds a set of lightweight UI elements to make interactions with LLMs more explicit and usable. 
At the top left, a persistent set of buttons supports navigation across substates, letting users move between note-taking, explicit LLM prompting, Insights, and back or confirm/cancel flows when needed. 
At the bottom right, a second set of controls provides the actions available within the current substate. 
This separation was designed to help users mentally distinguish transitions between substates from actions within a substate, while also supporting bimanual use: right-handed users can use the left hand for the top-left controls and the right hand with the pen for the bottom-right controls, while left-handed use reverses this arrangement.

Thinkink expands the types and formats of LLM Outputs. 
In addition to generation without request for Insights such as Socratic questions, it adds proofreading-oriented Insights; and in addition to generation with request for text-generation requests, it also supports image-generation requests, such as addressing users' direct image-generation requests or suggesting what to draw next. 
Thinkink also improves how these outputs are presented by supporting better-formatted text, including mathematical notation, so they are more useful in technical ideation tasks. 
In addition, both selected drawings and LLM generations can be revised through normal edits or LLM edits by annotation over text or image.

\begin{figure}
    \centering
    \vspace{-1mm}
    \includegraphics[width=1\linewidth]{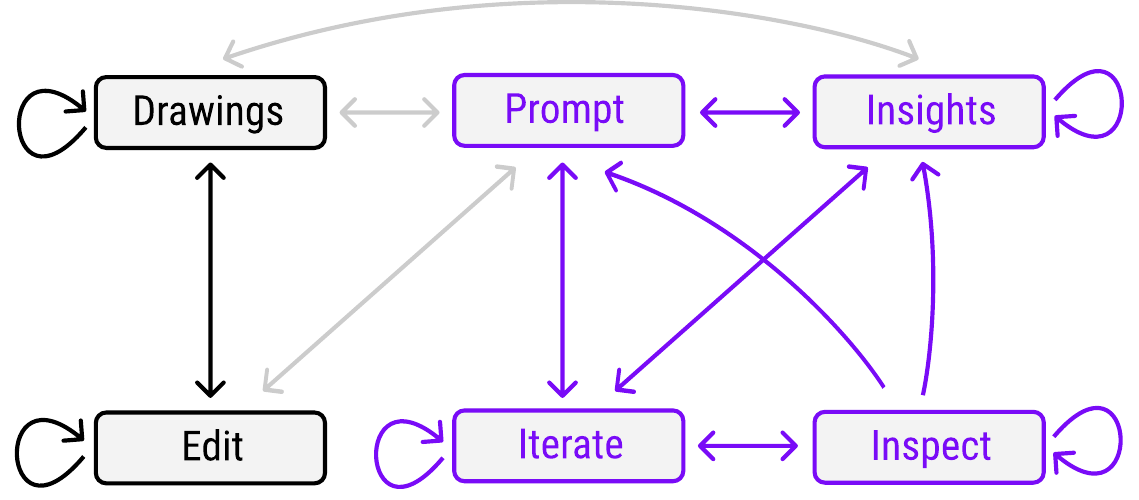}
    \caption{State Machine Interaction Design: Rectangles represent substates grouped into two superstates: ``Note-taking'' (``Drawings,'' ``Edit'') and ``LLM assistant'' (``Prompt,'' ``Insights,'' ``Iterate,'' ``Inspect''). Double-headed arrows show bidirectional transitions; single-headed arrows, unidirectional. Circular loops denote self-transitions (within-substate interactions). Edge colors encode transition type: black for transitions within ``Note-taking,'' indigo for transitions within ``LLM assistant,'' and cyan for cross-superstate transitions.}
    \label{fig:state-machine-overview}
    \vspace{-3mm}
\end{figure}

\subsection{Interaction Design using State Machine}

Thinkink organizes interactions as a substate machine.
This design choice is motivated by the first two UI and interaction challenges identified in our diagnostic study (\ie unclear boundaries between note-taking and LLM assistant, and limited support for error prevention, recovery, and user control) by making modes explicit, restricting each mode to a clear, non-overlapping set of interactions, and defining clear transitions for easier recovery and control.
The state machine includes \emph{two superstates}: \emph{Note-taking} and \emph{LLM assistant}, and \emph{six substates}: \emph{Drawings} and \emph{Edit} under \emph{Note-taking}, and \emph{Prompt}, \emph{Insights}, \emph{Iterate}, and \emph{Inspect} under \emph{LLM assistant}. Following Figure~\ref{fig:state-machine-overview}, we refer to between-substate transitions using labels such as \emph{Prompt->Iterate}, and to self-transitions as actions performed while in a given substate.

In \emph{Note-taking superstate}, \emph{\textbf{Drawings}} is the default substate for ordinary canvas work. While in Drawings, users can move existing drawings through ``dragging a drawing.'' Starting a new drawing with the drawing pen or selecting an existing drawing through ``tap-and-hold on a drawing'' triggers \emph{Drawings->Edit}, making that drawing the current focus of interaction.

While in \emph{\textbf{Edit}}, users can edit the selected drawing in two ways. 
First, they can perform normal editing using the drawing pen and the lasso eraser. 
Second, they can perform LLM editing using the LLM annotation pen and LLM annotation eraser to annotate the drawing and invoke LLM-supported revision of that note. 
In both editing modes, users can use the history bar to undo or redo revisions. 
Users can enter Edit through \emph{Drawings->Edit} or \emph{Prompt->Edit}. When they leave Edit, they transit to the previous substate.

In \emph{LLM assistant superstate}, \emph{\textbf{Prompt}} supports explicit prompting.
While in AA, the drawings on the canvas are treated as available context. 
Users can start an explicit request with the LLM prompting pen, which triggers \emph{Prompt->Iterate}. 
Users can also move existing drawings through ``dragging a drawing'' or open a drawing in Edit through ``tap-and-hold on a drawing,'' which triggers \emph{Prompt->Edit}. 
This makes explicit prompting a distinct interaction path rather than something inferred implicitly from general writing.

\emph{\textbf{Insights}} enables working with the Insight Nodes Layer. 
While in Insights, users inspect suggestions generated from the existing canvas without first authoring a request. 
Selecting a request node through ``tap-and-hold on a request'' triggers \emph{Insights->Iterate}, while selecting a generation-without-request insight node through ``tap on a generation'' triggers \emph{Insights->Inspect}. 
Users can also use the history bar to move across Insights sets and regenerate a new set; during regeneration, previous Insights are included in the prompt so the model is encouraged to produce different suggestions. 
Leaving this substate with the top-left navigation buttons returns users through \emph{Insights->Prompt}, preserving the distinction between Insights browsing and explicit prompting.

\emph{\textbf{Iterate}} is the main substate for refining an explicit LLM request and producing different generations from it. 
While in AR, users can continue writing and editing their requests using the LLM prompting pen, reposition requests via dragging, manually select which drawings serve as context using the lasso-based tool, and switch between generation types, such as text or image, which are automatically identified by the model.
As they refine the request or its context, new generations can be shown as responses. 
Users can also browse alternative generated responses while staying in AR using the history bar, which supports both revisiting prior alternatives and regenerating new ones. 
When regenerating, previous generations are included in the prompt so the model can produce new responses rather than repeat earlier ones. 
Selecting one of those generations through ``tap on a generation'' triggers \emph{Iterate->Inspect}. 
Leaving this substate with the top-left navigation buttons returns users through \emph{Iterate->Prompt} or \emph{Iterate->Insights}, depending on whether the request was reached from Prompt or Insights.

\emph{\textbf{Inspect}} is the inspection and decision substate for one LLM Output. 
While in AG, users can inspect the selected generation, edit it with the LLM annotation pen and LLM annotation eraser, move it on the canvas through ``dragging the generation,'' and use the history bar to undo or redo revisions. 
Back-navigation through the top-left buttons returns users through \emph{Inspect->Iterate} when the generation was reached from AR, or through \emph{Inspect->Insights} when it was reached directly from Insights. 
Confirming a generation is done by tapping the top-left confirm button, which accepts it into the workspace, turns it into a drawing, and returns to the previous active substate through \emph{Inspect->Prompt} or \emph{Inspect->Insights}. 
When \emph{Inspect->Insights} occurs, the system preserves the previously shown Insights. 
This allows users to work through one insight at a time, finalize it, and then return to continue with the remaining insights.

\section{Usage Study}

To address 
\emph{RQ3: How do users utilize Thinkink in practice?}
we conducted a usage study to understand how people use Thinkink for ideation. 
Our goal was to understand common usage patterns of such as tool in real-world scenarios.
Additionally, we highlight how changes in Thinkink, compared to the technical probe, address challenges identified in the diagnostic study, while also revealing new challenges encountered by participants.

\subsection{Method}
The method followed the diagnostic study, with two modifications: the interview questions were revised to focus on usage patterns rather than system diagnosis (see Appendix~\ref{apx:Usage_InterviewQuestions}), and additional survey questions were added to assess whether the challenges identified in the diagnostic study with the technical probe were addressed by Thinkink (see Appendix~\ref{apx:survey_questions_responses}). 
Regarding participants, we invited five participants back from the diagnostic study (P1–P5; \textit{M}=3, \textit{F}=2; age 26--37) and recruited five new participants (P7–P11; \textit{M}=2, \textit{F}=3; age 22--42), with a variety in major: quantum information and computation, HCI, computational arts/game design, fluid mechanics, reinforcement learning and AI ethics, vision science, statistics, entrepreneurship, health and aging, and biotechnology.
The cohort included 3 undergraduate students, 2 Master’s students, 4 PhD students, and 1 postdoctoral researcher.
Overall, participants reported moderate use of pen and paper for thinking and ideation (\textit{M}=3.50, \textit{SD}=0.97 on a 5-point scale), low use of digital inking tools (\textit{M}=2.60, \textit{SD}=0.70), and moderate to frequent use of LLMs for thinking and ideation (\textit{M}=3.90, \textit{SD}=0.74).
The full study materials are provided in Appendix~\ref{apx:Usage}, including examples of final artifacts (Appendix~\ref{apx:Usage_Artifacts}) and pictures captured during task performance (Appendix~\ref{apx:pictures_participants}).

\begin{figure}
    \centering
    \includegraphics[width=1\linewidth]{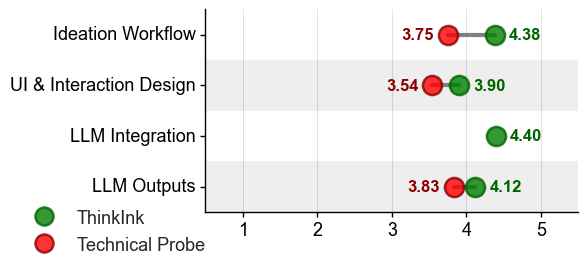}
    \vspace{-6mm}
    \caption{Average of participants ratings different prototypes}
    \label{fig:survey_summary}
    \vspace{-4mm}
\end{figure}

\subsection{Improvements and New Challenges}

The changes introduced in ThinkInk helped participants overcome several of the challenges identified in the diagnostic study, as reflected in interviews and survey responses (Figure~\ref{fig:survey_summary}; Appendix~\ref{apx:survey_questions_responses}). 
Participants also highlighted further improvements.

Regarding \textbf{UI and Interaction Design}, ThinkInk improved the previously unclear boundary between note-taking and LLM interaction by making modes more legible through in-situ visual cues. 
For example, P7 reported that distinguishing modes was \pquote{not really hard}{P7} because \pquote{this ink color appears on the corner}{P7}, suggesting that the interface better signaled when participants were writing notes versus engaging the model. 
ThinkInk also improved users’ ability to recover from errors and maintain control during interaction. After the model misunderstood part of a technical question, P1 was able to revise the prompt directly on the canvas rather than restart the task, explaining, \pquote{I erased it, and I made it simpler.}{P1} 
This indicates that the editable canvas supported lightweight error recovery and iterative refinement. 
A new issue, however, was that some participants felt a steep learning curve around modes and controls, with some users describing the experience as ``a little bit confused'' and warning that, without onboarding, they ``might be confused in the first place'': \pquote{I was a little bit confused}{P1}; \pquote{if you didn't give me instructions, I might be confused at the first}{P9}. 

Regarding \textbf{LLM Integration}, ThinkInk addressed the earlier inability to interact directly with LLMs by allowing participants to invoke the model when they were ready, instead of relying only on automatic suggestions. 
P5 highlighted this increased sense of agency: \pquote{So sometimes I knew what I wanted, so I right away asked it.}{P5} 
This suggests that ThinkInk restored more direct, user-initiated interaction with the LLM while keeping the exchange embedded in the shared note-taking workspace. 
Compared with conventional chat interfaces, this made LLM use feel more tightly coupled with participants’ ongoing work on the canvas. 
A new area for improvement was that participants wanted their prompts (requests) to persist on the canvas so they could feel a sense of ownership and review their finalizing, rather than having requests disappear after generation: \pquote{the request disappears}{P2}; \pquote{I want to keep track of my questions}{P1}. 

Finally, regarding \textbf{LLM Outputs}, ThinkInk partially addressed the earlier misalignment between LLM output formats and users’ needs by situating model responses within a canvas that also supported sketching and diagrammatic work, rather than constraining interaction to text-only chat. 
P1 explicitly contrasted ThinkInk with ChatGPT, noting that \pquote{The picture is much easier to draw, and it is impossible to do that in ChatGPT.}{P1} 
This suggests that ThinkInk better aligned LLM-supported work with the visual and diagrammatic nature of participants’ tasks. 
At the same time, output remained uneven: answers were sometimes too short, generic, or hallucinated \pquote{they were so short, sometimes they were very unrelated to what I asked}{P9} \pquote{it starts hallucinating}{P4}; and the model sometimes failed to follow instructions or produce accurate outputs in terms of the content \pquote{it generated me... a flow chart [instead of what the participant wanted]}{P3}.
Participants also sometimes felt that the model’s default context selection did not match their intent, leading to inaccurate responses. 
In such cases, they had to manually adjust the contexts themselves: \pquote{you have to go and deselect all the ones}{P4}; \pquote{it would not give me the response I wanted... so I had to undo (unselect) those contexts}{P9}.

\subsection{Usage Patterns}
We identified the following usage patters across the four design aspects.
As the data collected mirrored the diagnostic study, we used the same analysis process.

\subsubsection*{\textbf{Ideation Workflow: Note-taking paired with LLM assistant Supports Non-linear Iterative Ideation}} 
Participants often described ThinkInk as \pquote{one more partner}{P3} in ideation.
They used it for ideation through a spatial, non-linear workflow. 
They contrasted the 2D canvas with linear chat because it \pquote{gives you more freedom}{P1}, and P3 described serendipitous exploration~\cite{serendipity2017} through keeping multiple branches visible: \pquote{A is good, B is good, and C is good... I keep all of them... B and C, I will reach out to them later.}{P3}. 
We found that Thinkink can be used flexibly for varied by task (see Appendix~\ref{apx:usage_navigationflow} for usage patterns). 
For example, in a design creativity task, P2 generated alternative sketches and iterated on them, \pquote{I'm just sketching something... to see what kind of things AI (LLM) can do.}{P2}, while also noting that \pquote{as artists, we like to keep the history of the different iterations.}{P2}; 
P7 used the canvas to ideate plots as a branching map, \pquote{starting from the center... it leads to the next state of thinking.}{P7}. 
Table~\ref{apx:Usage_Participants_Choices} lists participants’ chosen tasks, and Appendix~\ref{apx:survey_questions_responses} shows they found Thinkink highly support ideation, when performing their tasks.
Note-taking also functioned as prompt grounding: P1 added written or drawn context because \pquote{it would be hard for this model to understand, that's why I was providing context}{P1}, and P2, when Insights were insufficient, turned back to Note-taking to supply more context and \pquote{highlight the part that's relevant, and then work from there.}{P2}.

\subsubsection*{\textbf{UI and Interaction Design: Cognitively Separating Note-Taking from the LLM assistant and Minimal UI Improves Usability}} 
Most participants described the interface as simple and minimal: \pquote{very simple... one person can understand and use}{P8}, while giving them the flexibility in using the features however they workflow requires (see Figure~\ref{apx:fig:usage_navigationflow}, including navigation and usage patterns).
They used the state machine interaction design as a understandable and memorable structure for available actions, although not all of them enacted the intended bimanual workflow of using the top-left controls with the left hand and in-substate interactions with the right. 
Some carried over familiar one-handed sketching habits (\pquote{when I sketch or create something in real life, I'm always just using my right hand... My left hand doesn't really do anything}{P2}) while most still referred to the left--right division as shaping their understanding of the interface because it \pquote{helps... build my mental model of what features does what}{P2} and made it \pquote{very helpful to have different icons... be on the other side}{P9}. 
Relatedly, some participants who used the right hand for all interactions described the hand movement from right-side to left-side buttons as creating a brief pause for thinking: \pquote{that time kind of works really well for when my right hand... moves there}{P2}. 
Participants also used lightweight UI elements in ways that reduced interaction overhead: output-type, including text and image, selection reduced prompt-writing burden (\pquote{I don't need to always specify my request... generate me an image}{P3}), and iconography helped them distinguish LLM functions from ordinary note-taking (\pquote{easy to identify... the AI (LLM) versus normal note tweaking}{P11}).

\subsubsection*{\textbf{LLM Integration: While Insights Used for Discovery, Explicit Prompts used for Follow-Up}} 
Participants used ThinkInk's Prompt mode in conjunction with note-making and steering. 
Notably, P8 used Prompt multimodally, effectively turning written text and even shapes into prompts, while emphasizing that successful interpretation depended on grounding the model in the current context, \pquote{So that's the reason I was making sure that the context is right}{P8}. 
They also used Insights to surface overlooked ideas and context-aware ideation: \pquote{it was giving me questions that I didn't know about.}{P10}; \pquote{knows, like, what all the stuff that, I had written. Based on that, it was providing me things that I could, like, ask}{P10}.
P8 also used Insights as an initial probe, but when those suggestions surfaced a different avenue, they returned to Prompt to re-steer, again foregrounding context management: \pquote{So that's the reason I was making sure that the context is right}{P8}. 
They sometimes confirmed one insight and switched to the Prompt mode to explore it further: \pquote{I would pick one of them that I like and continue with the AI querying mode to ask more questions about that insight}{P9}.

\subsubsection*{\textbf{LLM Outputs: Context Management is Necessary, But it Enables Shaping, Comparing, and Refining LLM responses}} 
Participants managed LLM Outputs on the spatial canvas differently from chat, noting that responses remained traceable rather than ``all linear'' \pquote{it's all linear.}{P3}. 
Across Prompt and Insights, they refined outputs through regeneration and history, using iteration to recover from wrong answers or move the model toward a new generation: \pquote{when you regenerate, the new one is correct.}{P1}; \pquote{try again ... check if there's new things.}{P7}. 
Context selection and manipulation were used to test different assumptions, specially in a quantum computing task done by P1 or a design creativity task by P2.
Also, it was used to adjust relevance and make the system's focus more explicit by revealing the context that is used to generate response \pquote{it's pointing out the focus (context)}{P7}, allowing users to remove distracting material \pquote{can you just deselect these contexts?}{P1}. 
\pquote{LLM editing was used mainly during convergence}, when users knew what to change \pquote{the LLM edit feature work better.}{P2} and annotation afforded control \pquote{giving a good control of the context.}{P7}.

\section{Discussion}

Ink-native interaction with LLMs could enable building new tools across domains such as mathematics, physics, and graphic design, as well as in scenarios including ideation, collaboration, and AR/VR. 
This paper represents a stepping stone toward realizing ink-native interactions with LLMs, an area that remains largely underexplored.
Although our studies focused on a specific scenario and domains, and involved a limited number of participants, they provide useful insights for future research. 
Across our formative, diagnostic, and usage studies, we demonstrate how to build a foundation for transforming unstructured drawings on a 2D spatial canvas into structured representations that is more understandable for LLMs.
Our findings further suggest that designing ink-native LLM-powered tools for ideation is not a matter of identifying a single optimal interface, but rather of balancing competing design goals across different users and tasks, focusing on different design aspects.

\subsubsection*{\textbf{Ideation Workflow: Linear vs. Non-linear Progression}}
The formative study showed that externalization is non-linear and often begins with incomplete marks, words, or structures. 
The diagnostic and usage studies then showed why this matters: participants used the 2D canvas to branch, keep alternatives visible, and ground prompts in drawings. 
However, not all users benefited from the same degree of openness. Some valued free-form exploration, while others wanted clearer structure. 
This suggests a key design choice for future systems: whether to privilege open spatial exploration or provide more linear scaffolds, summaries, or templates. 
The right balance will likely differ for different tasks, such as design creativity, technical reasoning, and everyday planning, as well as for users who naturally work in organized versus free-form ways.

\subsubsection*{\textbf{UI and Interaction Design: Constrained vs. Unconstrained Interactions}}
The studies consistently supported the value of a minimal interface design. 
Participants appreciated low visual clutter and pen-first interaction, which aligned with the immediacy of paper. 
The diagnostic study also showed that minimalism can become ambiguity when modes, system attention, and available actions are unclear. 
Thinkink's state-machine design improved legibility and control, but the usage study still revealed a learning curve. 
This requires a design decision: preserving a low-friction interface versus adding explicit constraints, cues, and onboarding. 
More constrained interactions may especially benefit users with less experience using digital inking tools, while expert users may prefer fewer guardrails and more interaction choices.
Adaptive disclosure may therefore be promising: simple defaults for novices, with more flexible interactions revealed as users gain confidence.

\subsubsection*{\textbf{LLM Integration: User vs. LLM Agency}}
Participants across studies wanted LLMs to act as an ideation partner embedded in the canvas, not as a separate chatbot. 
They valued Insights that expanded thinking, but also wanted ways to explicitly ask the model, control when it acted, and inspect or adjust context. 
Together, these findings suggest that Insights and Prompt should be treated as separate but interoperable design dimensions rather than a single continuum. 
Different tasks may require varying levels of LLMs initiative: tasks where users are uncertain about their goals, such as everyday decision-making, can benefit from more Insights, whereas tasks where users have clearer objectives, such as specialized work like math problem-solving, may require more explicit prompting.

\subsubsection*{\textbf{LLM Outputs: Generic vs. Task-specific Content and Representations}}
Exploratory questions were useful for broad ideation, but generic text-only outputs were insufficient for technical and creative work. 
Participants wanted diagrams, better formatting, mathematical notation, and editable visuals. 
Thus, future ink-native LLM-powered tools should decide whether to optimize for general-purpose support or for domain-tuned outputs matched to target tasks such as mathematics, design creativity, or scientific reasoning.

\section{Conclusion}

Handwritten and sketch-based ideation helps people externalize and explore ideas, yet current LLM interfaces remain largely chat-based and poorly suited to spatial ink workflows. 
We presented \textit{Thinkink}, a LLM-assisted ideation tool that supports ink-based prompting and on-canvas LLM responses, including a semantic-tree representation of the workspace and an explicit state-machine interaction design. 
Through a user-centered, three-stage process, a formative focus group ($n=12$), a diagnostic study of a technical probe ($n=6$), and a usage study of \textit{Thinkink} ($n=10$), we identified and refined four core design aspects: ``Ideation Workflow,'' ``UI and Interaction Design,'' ``LLM Integration,'' and ``LLM Outputs,''. Our results suggest that future ink-native LLM-assisted tools should treat these aspects as balances to tune for target users and tasks. It should explore adaptive interfaces, richer multimodal outputs, and domain-specific versions for settings such as math, design creativity, and everyday decision making.

%% reference section
\bibliographystyle{ACM-Reference-Format}
\bibliography{_references}

\onecolumn
\appendix
\clearpage

\renewcommand\thefigure{\thesection.\arabic{figure}}
\renewcommand\thetable{\thesection.\arabic{table}}
\setcounter{figure}{0}
\setcounter{table}{0}
\section{Formative Study}

\subsection{Question Categories and Examples}
\label{apx:Formative_Question_Categories}

\begin{table}[h]
\centering
\small
\rowcolors{2}{gray!15}{white}
\begin{tabularx}{\linewidth}{p{3.5cm} X}
\toprule
\textbf{Category} & \textbf{Description and Example} \\
\midrule

Brainstorming Questions & Encourage creative thinking, idea generation, and innovation by exploring a wide range of possibilities. \newline
\textit{Example:} ``I'm designing a system for a research paper. What would be a suitable name for it?'' \\

Personal Questions & Ask about an individual's experiences, opinions, or feelings. \newline
\textit{Example:} ``What motivates me to start my day?'' \\

Open-Ended Questions & Encourage detailed responses and cannot be answered with a simple ``yes'' or ``no.'' \newline
\textit{Example:} ``How do technological advancements shape the way we communicate?'' \\

Thought-Provoking Questions & Stimulate deeper reflection and contemplation, often challenging assumptions. \newline
\textit{Example:} ``What does success mean?'' \\

Cause-and-Effect Questions & Examine relationships between actions or events and their outcomes. \newline
\textit{Example:} ``How does employee engagement influence productivity?'' \\

Comparative Questions & Ask for comparisons between two or more ideas or experiences. \newline
\textit{Example:} ``Which is more fulfilling: working independently or in a team, and why?'' \\

Exploratory Questions & Investigate new or unknown areas to gain understanding. \newline
\textit{Example:} ``What opportunities could emerge from this new technology?'' \\

Planning Questions & Focus on creating a roadmap or strategy for future actions. \newline
\textit{Example:} ``How would you structure a six-month plan to achieve this goal?'' \\

Problem-Solving Questions & Aim to identify and resolve issues or challenges. \newline
\textit{Example:} ``How can we fix issues in the current system?'' \\

Decision-Making Questions & Involve choosing between options or prioritizing actions. \newline
\textit{Example:} ``If you had to focus on one project this quarter, which would it be and why?'' \\

Critical Thinking Questions & Require analyzing and evaluating information to form judgments. \newline
\textit{Example:} ``What factors contribute to procrastination?'' \\

Career Questions & Focus on professional development and career paths. \newline
\textit{Example:} ``Did I choose the right career path?'' \\

\bottomrule
\end{tabularx}
\vspace{1mm}
\caption{Question categories used in the formative study.}
\end{table}

\subsection{Participants Choices for the tasks}
\label{apx:Formative_Participants_Choices}

\begin{table*}[h]
\centering
\small
\rowcolors{2}{gray!15}{white}
\begin{tabularx}{\textwidth}{p{2.5cm} p{2.5cm} p{3cm} X}
\toprule
\textbf{Participant ID} & \textbf{Task Type} & \textbf{Medium} & \textbf{Selected Question} \\
\midrule

1047 & Individual & Pen \& Paper & Why do plans get delayed, and what factors cause procrastination? \\

1153 & Individual & Pen \& Paper & Did I choose the right career path? \\

1253 & Individual & Pen \& Paper & - \\

1300 & Individual & Pen \& Paper & How would I structure a plan to get a faculty job immediately after I graduate? \\

1473 & Individual & Pen \& Paper & What motivates me to purchase online courses? \\

1540 & Individual & Digital Inking & - \\

1571 & Individual & Figma & How to plan my project to meet the next deadline? \\

1766 & Individual & Pen \& Paper & How do I add a notification feature to my current system? \\

1820 & Individual & Digital Inking & How could personal habits/behaviors gradually change a system or interface to better serve users over time? \\

1842 & Individual & Pen \& Paper & If you had to focus on only one project this quarter, which would it be and why? \\

1902 & Individual & Pen \& Paper & What makes a good presentation? \\

-- & Individual & Pen \& Paper & -- \\

-- & Collaborative & Pen \& Paper & How to not be overwhelmed? \\

-- & Collaborative & Pen \& Paper & -- \\

-- & Collaborative & Pen \& Paper & -- \\

\bottomrule
\end{tabularx}
\vspace{1mm}
\caption{Participants' selected tasks}
\end{table*}

\subsection{Discussion Points}
\label{apx:Formative_Discussion_Points}

\begin{itemize}
    \item How did the process of externalizing evidence influence your understanding and the quality of your answer to the question? Please explain why it had an impact, or why it did not.

    \item Did visualizing the evidence lead you to discover any new connections or insights that you hadn't previously considered? If so, please elaborate.

    \item In what other types of questions or scenarios do you think externalizing evidence would be beneficial? Conversely, in what situations might it not be helpful?

    \item How easy or difficult was it for you to externalize the evidence, and what factors influenced that experience?

    \item Why do you think the tool you used is or isn’t suitable for this task?

    \item In what aspects did you collaborate, and what challenges did you encounter (if any)?

    \item Why did you choose or not choose to use ChatGPT for this task? if you chose, for what? What are the scenarios that ChatGPT might help?
\end{itemize}

\subsection{Artifacts}
\label{apx:Formative_Artifacts}

\begin{figure}[h]
    \centering
    \includegraphics[width=1\linewidth]{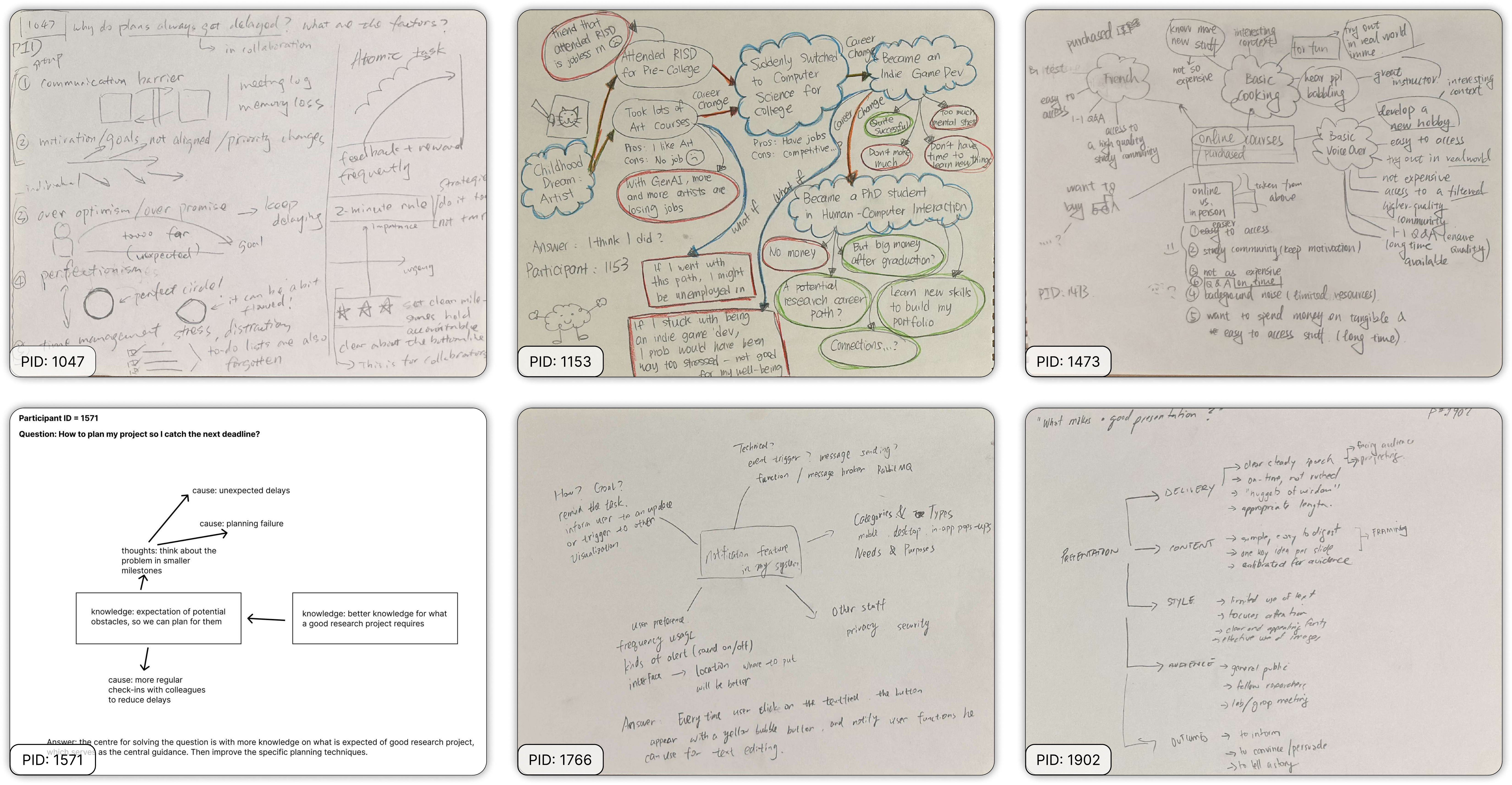}
    \caption{Examples of artifacts made by participants during the formative study}
    \label{fig:Formative_Artifacts}
\end{figure}

\section{Technical Probe}

\subsection{Prompt for Base Semantic Tree Generation}
\label{apx:base-semantic-tree-generation}

\noindent
\begin{tcolorbox}[
    colback=gray!10,
    colframe=gray!50!black,
    fontupper=\ttfamily\footnotesize,
    breakable,
    title=System instruction for generating new code \\ (promptType: gen\_code)
]

Analyze this hand-drawn thinking canvas and build its base semantic tree.

Build the semantic tree for this hand-drawn canvas.

The tree must contain ONLY "drawing" and "concept" nodes.  
No other node types are allowed.

For each drawing, analyze and include in the drawing node:

- drawingType (text/shape)

- text: extracted text content (if drawingType is text)

- desc: description considering canvas context without mentioning other drawings  
  (if drawingType is shape; maximum 5 words); as it is hand-drawn, don't invent  
  overly specific details. If the shape contains text, include the text as part  
  of the description (e.g. "Rectangle with text 'Hello'").

NODE TYPES (only these two):

- "drawing" = individual sketch/text on the canvas (one per drawing)

- "concept" = meaningful group of 2+ related items (drawings or concepts)

RELATIONSHIP \& GROUPING RULES — READ CAREFULLY:

- Only group drawings under a shared "concept" node if they have a CLEAR,  
  SPECIFIC, and MEANINGFUL semantic relationship.

- DO NOT group drawings together just because they are on the same canvas,  
  are spatially near each other, or are both hand-drawn items.

- DO NOT create vague concept labels like "Ideas", "Notes", "Thoughts",  
  "Canvas Items", "Topics", "Concepts", or "Main Ideas".

- If drawings cover DIFFERENT, UNRELATED topics, they MUST remain as separate  
  root nodes.

- Prefer FLAT structures with multiple independent roots over a single deeply  
  nested tree.

- A concept node is only justified when removing it would lose meaningful  
  structural information about how its children relate to each other.

STRUCTURAL RULES:

- Total drawing nodes must EXACTLY equal total individual drawings provided

- Concept = parent with 2+ children (drawing or concept)

- Multiple roots are expected and preferred when drawings are unrelated

- No duplication of similar nodes

Node fields:

drawing: id, nodeType, drawingId (string index), drawingType, text/desc, children  

concept: id, nodeType, conceptLabel, children

Return ONLY valid JSON, no additional text or explanations.

Return JSON format:

{

  "semanticTree": [...]
  
}

CRITICAL - DRAWING INDEX MAPPING:

- There are EXACTLY N drawings (indices 0 to N-1)

- The "semanticTree" MUST contain a drawing node with drawingId for ALL N drawings using their indices (0-N-1)

- DO NOT skip any drawing index

- Use drawingId as an INDEX (0, 1, 2, ...) not as a string ID

\end{tcolorbox}

\subsection{Prompt for the Insight Nodes Layer Generation}
\label{apx:proactive-node-generation}

\noindent
\begin{tcolorbox}[
    colback=gray!10,
    colframe=gray!50!black,
    fontupper=\ttfamily\footnotesize,
    breakable,
    title=System instruction for generating new code \\ (promptType: gen\_code)
]

Given this existing semantic tree of a hand-drawn thinking canvas:
<treeJson>

<DRAWING DIMENSIONS section if available>

Generate AI insights nodes and attach them as children of the appropriate  

drawing nodes or concept nodes.

Node types to generate:

"text\_generation\_request":  

a concise request (5–15 words) for LLM-generated text information related  

to the parent node's content

"generation\_without\_request": insight output with one of these subtypes:

- insightSubtype="socratic": a thought-provoking Socratic question  
  (5–15 words) that deepens thinking about the parent node

- insightSubtype="suggest\_next\_drawing": a concise suggestion (1–4 words)  
  for a specific next drawing element that the user might add near the  
  parent node

RULES:

- These nodes can be children of drawing nodes OR concept nodes

- Multiple insight nodes of different types can be attached to the same parent

- No duplication of similar nodes

- A node should be attached to the parent where the content is most relevant

- Generate insight nodes only when they meaningfully support ideation

- generation\_without\_request nodes may use ONLY these two subtypes:  
  "socratic" and "suggest\_next\_drawing"

- For suggest\_next\_drawing, output the name of a specific drawing element,  
  NOT an instruction sentence. Examples: "arrow", "circle", "label",  
  "branch", "example box"

- For responsePosition, choose a location based on available space and  
  meaningful placement relative to the parent drawing or concept

Node fields:

text\_generation\_request: id, nodeType, requestPrompt, responsePosition,  
fontSize, suggestedWidth, suggestedHeight, children

generation\_without\_request: id, nodeType, insightSubtype  
("socratic"|"suggest\_next\_drawing"), requestPrompt, responsePosition,  
fontSize, suggestedWidth, suggestedHeight, children

fontSize / suggestedWidth / suggestedHeight rules:

- fontSize: integer (12–220), estimated from parent drawing scale

- suggestedWidth: max(parent drawing width × 2, fontSize × 12), minimum 350px

- suggestedHeight: roughly parent drawing height, minimum 50px

Return the full modified semantic tree:

{

  "semanticTree": [...]
  
}

\end{tcolorbox}

\subsection{Prompt for Text Response Generation}
\label{apx:text-response-generation}

\noindent
\begin{tcolorbox}[
    colback=gray!10,
    colframe=gray!50!black,
    fontupper=\ttfamily\footnotesize,
    breakable,
    title=System instruction for generating new code \\ (promptType: gen\_code)
]

The user has drawn something and is asking:

"<requestPrompt>"

CONTEXT:

<context item 1>  

<context item 2>  

...

Respond in relation to the context.

Provide a brief, direct answer (max 25 words).

\end{tcolorbox}

\section{Diagnostic Study}
\label{apx:Diagnostic}

\subsection{Interview Questions}
\label{apx:Diagnostic_InterviewQuestions}

\textbf{General}
\begin{enumerate}
    \item What are your general thoughts and insights?
    \item How would you compare this experience to trying to solve this problem using a standard chatbot (like ChatGPT)?
    \item How would you compare this to solving the problem on a physical whiteboard or paper without AI?
\end{enumerate}

\textbf{Four aspects: Ideation Workflows, UI and Interaction Design, AI Integration, and AI Outputs}:
\begin{enumerate}
    \item What challenges did you face when doing the tasks?
    \item You performed two tasks. Did you find yourself using the tool differently between the two? How so?
    \item How can the tool help you perform your task? What feature is missing?
    \item Were there moments where you wanted to express an idea to the AI but felt the sketching/writing tools limited you? If so, what were these moments?
    \item Did you encounter limitations when trying to convey ideas to the AI using the sketches themselves? If so, what were you trying to communicate?
    \item How did you feel about using a 2D space to communicate with AI?
    \item What was your experience using a 2D canvas to perform the tasks? Did it feel like using pen and paper or not? Why?
    \item What was your experience with the merging feature?
    \item When the AI responded, did the content match what you were looking for? If not, what did you expect to see?
    \item When the AI responded, did the presentation/format match what you were looking for? If not, what did you expect to see?
    \item How did you feel about the visual presentation of the AI's answers? Did they clutter your space, or did they fit into your flow?
    \item Can you point to a specific moment where the AI's response triggered a ``lightbulb moment'' or a shift in your reasoning?
\end{enumerate}

\textbf{Closing}
\begin{enumerate}
    \item Do you have any further thoughts or suggestions for enhancing the tool?
\end{enumerate}

\subsection{Participants Choices for the tasks}
\label{apx:Diagnostic_Participants_Choices}

\begin{table*}[h]
\centering
\small
\begin{tabularx}{\textwidth}{l X X}
\toprule
\textbf{PID} & \textbf{Task 1: Chosen} & \textbf{Task 2: Self-defined} \\
\midrule

\rowcolor{gray!15}
P1 & How can a professor increase efficiency and research output without constantly pushing group members? & Does an ``area law'' exist for tripartite GHZ entanglement? \\

P2 & Which is more fulfilling: working independently or in a team, and why? & The ATC horror game. \\

\rowcolor{gray!15}
P3 & How can I maintain a healthy lifestyle? & In computer science education, how should AI tools be designed to support students’ reasoning and learning? \\

P4 & How can I plan my day and reduce my social media usage? & What is the best assumption for a specific fluid problem? \\

\rowcolor{gray!15}
P5 & How do philosophers draw conclusions from a given idea? & How can I integrate social learning into an ethical reinforcement learning process? \\

P6 & Which does an LLM suggest: traveling by official transportation services or using rideshare? & How can I simulate the polarization profile after passing through optical elements? \\

\bottomrule
\end{tabularx}
\vspace{1mm}
\caption{Participants' selected tasks organized by category}
\end{table*}

\clearpage

\subsection{Artifacts}
\label{apx:Diagnostic_Artifacts}

\begin{figure}[h]
    \centering
    \includegraphics[width=1\linewidth]{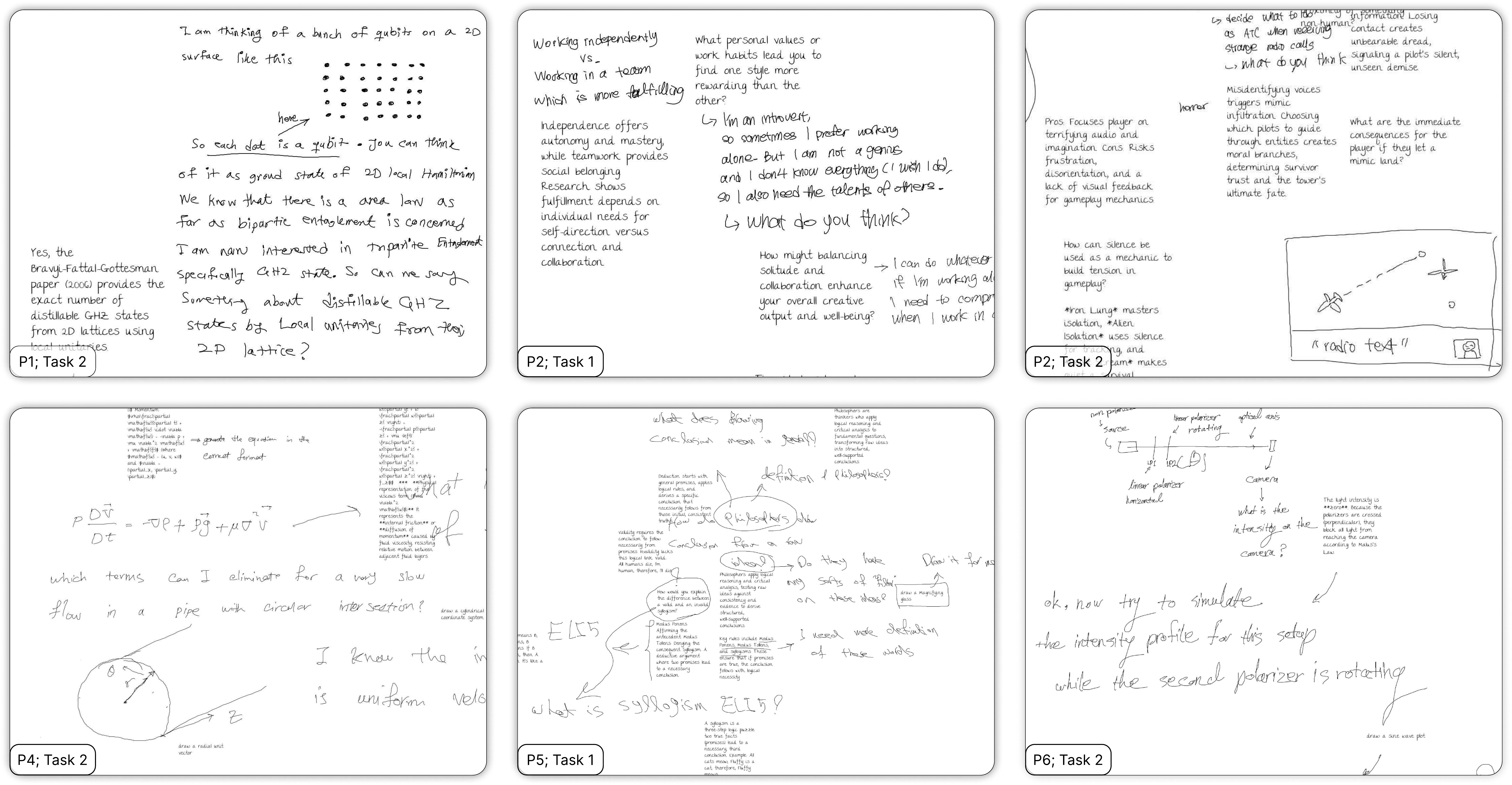}
    \caption{Examples of artifacts made by participants during the diagnostic study}
    \label{fig:Formative_Diagnostic}
\end{figure}

\clearpage

\section{Usage Study}
\label{apx:Usage}

\subsection{Interview Questions}
\label{apx:Usage_InterviewQuestions}

\paragraph{General}
\begin{enumerate}
    \item What are your general thoughts and insights?
    \item How would you compare this experience to trying to solve this problem using a standard chatbot (like ChatGPT)?
    \item How would you compare this to solving the problem on a physical whiteboard or paper \textit{without} AI?
\end{enumerate}

\textbf{Four aspects: Ideation Workflows, UI and Interaction Design, AI Integration, and AI Outputs}:
\begin{enumerate}
    \item How did the tool support or hinder your ability to connect ideas and build a conclusion today? What worked well, and what felt unnatural?
    \item Can you describe a specific moment where you reached a mental block or knowledge gap, and how the tool helped (or failed to help) you move forward?
    \item Did the spatial layout of the canvas change the way you reasoned through your problem compared to a standard vertical document? How?
    \item How did the AI change what you wrote/drew or how you organized the canvas?
    \item How did having the AI responses \textit{directly on the 2D canvas} (rather than in a linear chat) affect how you synthesized information?
    \item Did the tool feel like a unified workspace, or did you feel like you were switching between ``thinking mode'' and ``AI mode''?
    \item How do you feel about the interactions? Was it easy to get the most of the tool using the interactions it required?
    \item At what moments did you decide to use (or not use) the AI? What triggered those decisions?
    \item In what scenarios did you prefer Ask AI over AI insights, or vice versa?
    \item How transparent did the AI feel to you in terms of which parts of the canvas it was reading or using to generate its responses?
    \item Did you feel you had enough manual control \textit{to change} information the AI was looking at before it gave you an answer?
    \item How did this level of control affect your trust in the accuracy and relevance of the AI's responses?
    \item How did you manage the sheer amount of information the AI gave you? Did you use the Generation History?
    \item How was your experience modifying or editing the AI's generated text/images?
\end{enumerate}

\textbf{Challenges}:
\begin{enumerate}
    \item Were there any moments where the tool did something you didn't expect? If so, how easily were you able to undo it or recover your flow?
    \item Did the AI insights ever feel intrusive or interrupt your train of thought?
    \item Did you ever feel anxious about ``losing'' an AI response like you might in a continuous chat thread? How did the tool alleviate or worsen this?
\end{enumerate}

\textbf{}{Returning Participants vs.\ New Participants}
\begin{itemize}
    \item \textit{For Returning Participants:} You tested an earlier version of the tool. What stood out to you as the biggest improvement today? Did any of the changes introduce \textit{new} problems for your workflow?
    \item \textit{For New Participants:} If you were to adopt this tool for your domain tomorrow, what is the very first feature you would use and what would you change?
\end{itemize}

\paragraph{Closing}
\begin{enumerate}
    \item Do you have any further thoughts and insights? Additionally, do you have any suggestions for enhancing the tool?
\end{enumerate}

\subsection{Participants Choices for the tasks}
\label{apx:Usage_Participants_Choices}

\begin{table*}[h]
\centering
\small
\rowcolors{2}{gray!15}{white}
\begin{tabularx}{\textwidth}{l X X}
\toprule
\textbf{PID} & \textbf{Task 1: Chosen} & \textbf{Task 2: Self-defined} \\
\midrule

P1 & What motivates me to learn about history & The capacity of wiretap channel \\
P2 & Decide about weapon purchase & Brainstorming my next painting \\
P3 & What does success mean? & How to teach people to use a debugger \\
P4 & How would you structure a six-month plan to achieve my goal? & Dependency flow inside the pipe to geometric parameters \\
P5 & Gym planning & Understanding semi MDP \\
P7 & How would you structure a six-month plan to achieve this goal? & Ways to do detrending \\
P8 & If you had to focus on only one project this quarter, which would it be and why? & Learn about how entrepreneurs contribute to sustainability \\
P9 & What motivates me to continue my career? & Brainstorming my PhD thesis topic (health) \\
P10 & How can we fix the issues we're facing with food shortages? & How to use AI for sustainable development? \\
P11 & Where do you see yourself in five years professionally? & Drawing a plasmid to enter into E.\ coli host, ensuring I have all features needed \\

\bottomrule
\end{tabularx}
\vspace{1mm}
\caption{Participants' selected tasks. Returning participants were P1--P5; new participants were P7--P11. Returning participant P6 did not join this study.}
\end{table*}

\clearpage

\subsection{Artifacts}
\label{apx:Usage_Artifacts}

\begin{figure}[h]
    \centering
    \includegraphics[width=1\linewidth]{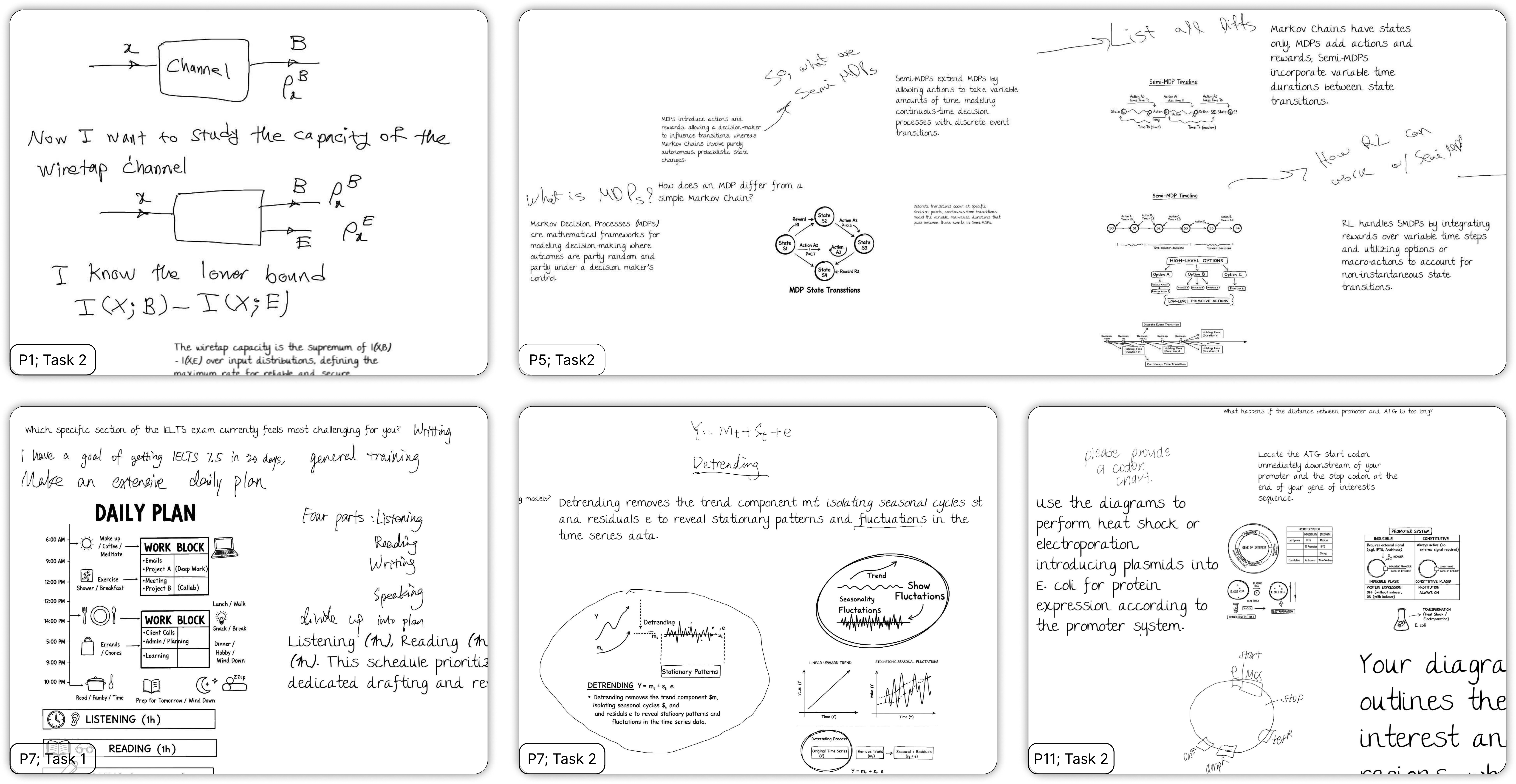}
    \caption{Examples of artifacts made by participants during the usage study}
    \label{fig:Formative_Usage}
\end{figure}

\subsection{Pictures of Participants Using Thinkink}
\label{apx:pictures_participants}

\begin{figure}[h]
    \centering
    \includegraphics[width=1\linewidth]{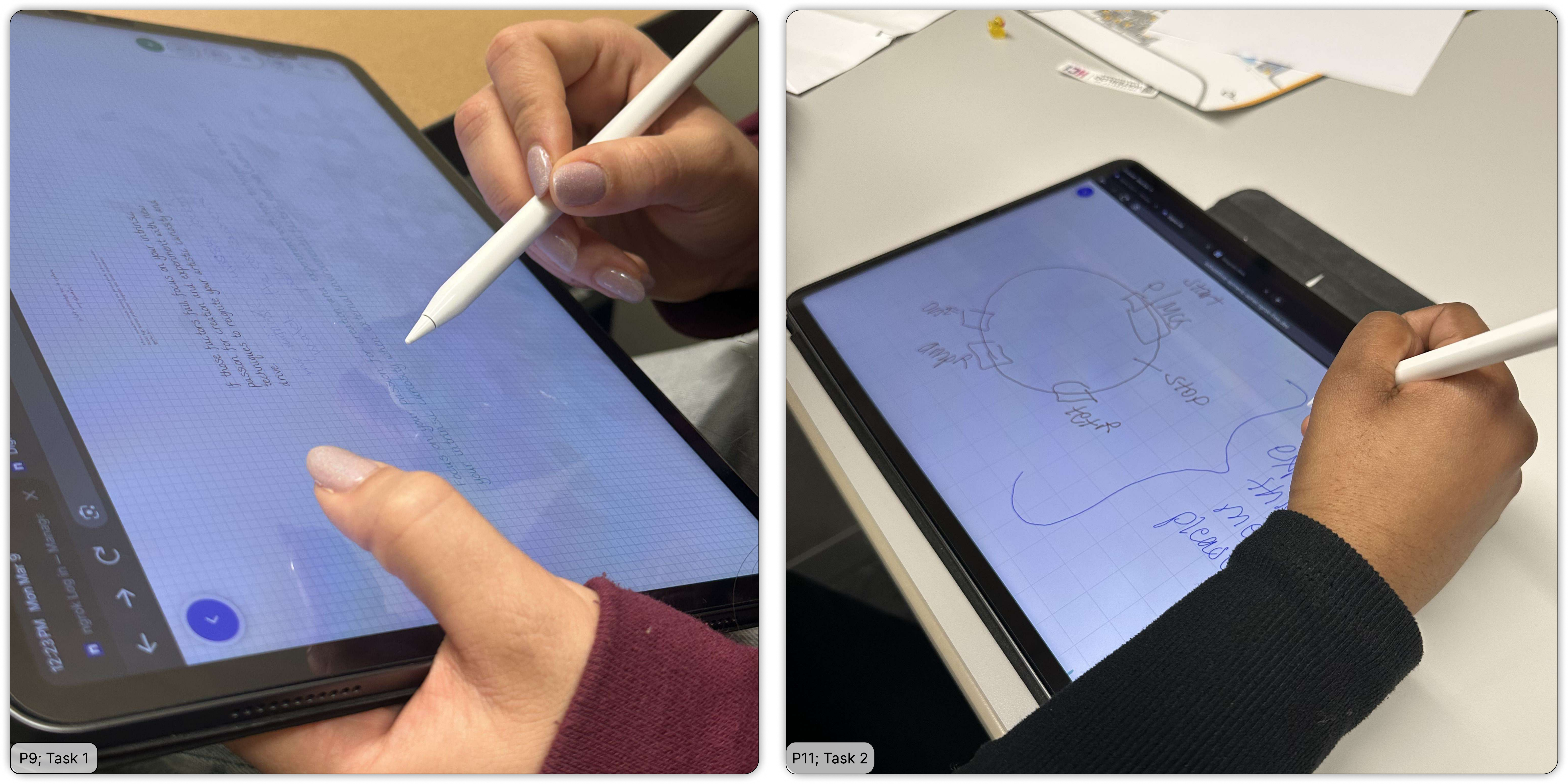}
    \caption{Pictures of Participants Using Thinkink}
    \label{fig:pictures_participants}
\end{figure}

\subsection{Pictures of Participants Using Thinkink}
\label{apx:pictures_participants}

\begin{figure}[h]
    \centering
    \includegraphics[width=1\linewidth]{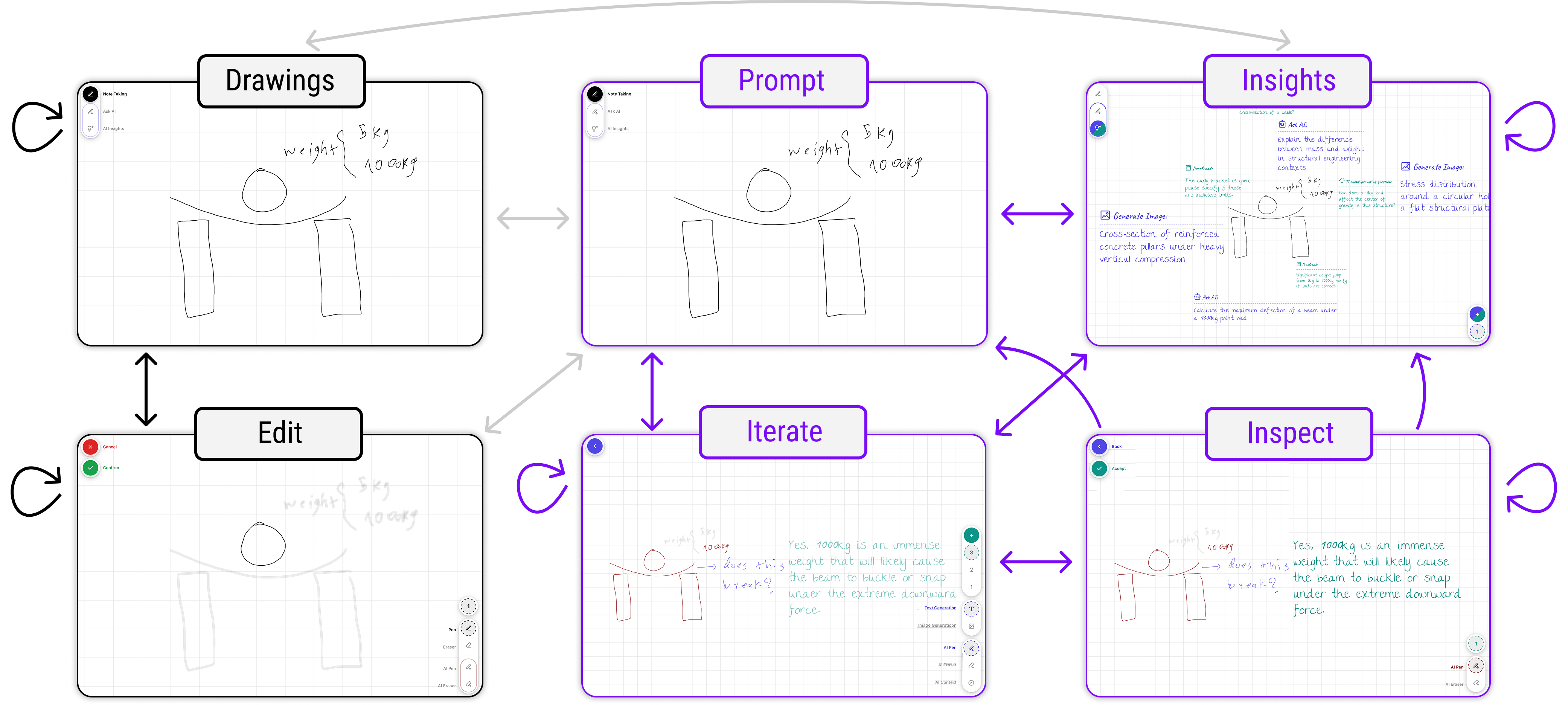}
    \caption{Detailed version of State Machine Interaction Design, incluging Thinkink Screenshots. Rectangles represent substates grouped into two superstates: “Note-taking” (“Drawings,” “Edit”) and “LLM assistant” (“Prompt,” “Insights,” “Iterate,” “Inspect”). Double-headed arrows indicate bidirectional transitions, while single-headed arrows denote unidirectional transitions. Circular loops represent self-transitions (within-substate interactions). Edge colors encode transition types: black for transitions within “Note-taking,” indigo for transitions within “LLM assistant,” and cyan for cross-superstate transitions. This detailed version augments this state machine with accompanying screenshots, providing concrete examples of each substate and illustrating how transitions are manifested in the interface.}
    \label{fig:screenshots-statemachine}
\end{figure}

\clearpage

\subsection{State Machine Interaction Design Usage Patterns}
\label{apx:usage_navigationflow}

\begin{figure*}[h]
    \centering
    \includegraphics[width=1\linewidth]{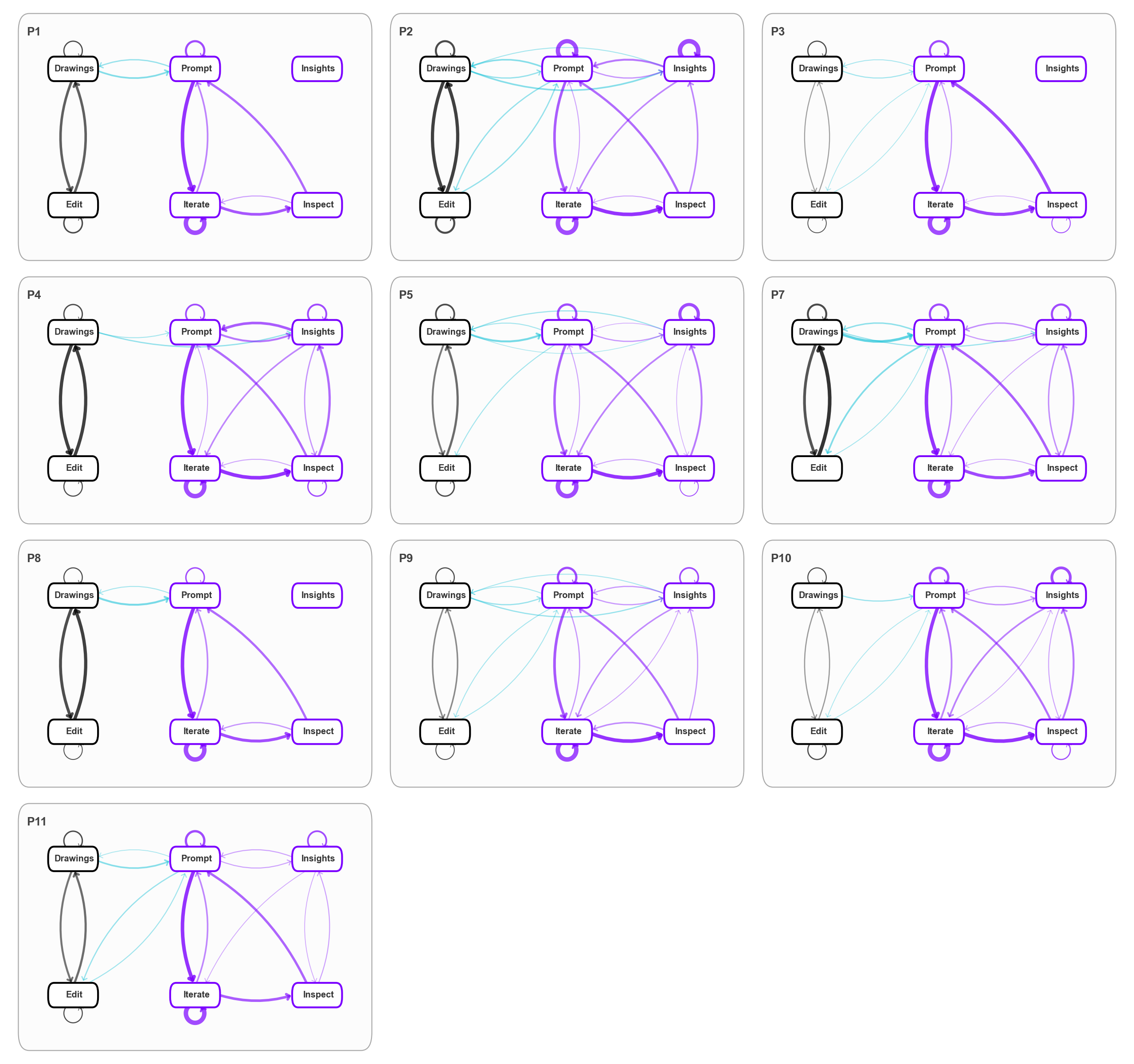}
    \caption{Per-participant state diagrams demonstrate how each participant navigated among the six substates of ThinkInk, with edge thickness indicating transition frequency and self-loops indicating repeated interaction within a substate. Arrow colors encode transition type: black edges indicate transitions within "Note-taking" substates, indigo edges indicate transitions within "AI assistant" substates, and cyan edges indicate transitions between the two superstates. The figure highlights substantial flexibility in use patterns: some participants, including P2, P4, P7, and P8, highly used the "Note-taking" superstate, including "Drawings" and "Edit" substates, whereas others, including P3, P9, and P10, concentrated much more of their interaction in the "AI assistant" superstate than the "Note-taking" superstate. "AI Insights" was used by all participants except P3, P8, and P10, further underscoring that the interface afforded multiple viable ways of working rather than steering users into a single prescribed workflow. At the same time, "Iterate" emerged as the most consistently prominent substate across participants, suggesting that refinement of generations was a central part of interaction with the system.}
    \label{apx:fig:usage_navigationflow}
\end{figure*}

\clearpage

\section{Diagnostic and Usage Studies survey questions' responses}
\label{apx:survey_questions_responses}

\begin{table*}[h]
\centering
\footnotesize
\renewcommand{\arraystretch}{1.15}

\definecolor{diaggray}{gray}{0.92}
\definecolor{usagegray}{gray}{0.85}

\begin{tabularx}{\textwidth}{l X 
>{\columncolor{diaggray}}c >{\columncolor{diaggray}}c 
>{\columncolor{usagegray}}c >{\columncolor{usagegray}}c}
\toprule
& & \multicolumn{2}{c}{\textbf{Diagnostic study (technical probe)} ($n{=}6$)} & \multicolumn{2}{c}{\textbf{Usage study (Thinkink)} ($n{=}10$)} \\
\cmidrule(lr){3-4}\cmidrule(lr){5-6}
\textbf{Code} & \textbf{Question} & \textbf{Median (IQR)} & \textbf{Mean (SD)} & \textbf{Median (IQR)} & \textbf{Mean (SD)} \\
\midrule

\rowcolor{black}
\multicolumn{2}{l}{\textcolor{white}{\textbf{Ideation Support and User Workflows}}} 
& & \textcolor{white}{3.75} 
& & \textcolor{white}{\textbf{4.38}} \\
\midrule
Q1.1 & The tool helped me generate new explanations, hypotheses, or information I had not considered.
& 4.00 (2.50--4.00) & 3.50 (1.22)
& 4.00 (4.00--4.75) & \textbf{4.30 (0.48)} \\
Q1.2 & I felt satisfied with the conclusion I reached at the end of the session.
& 4.00 (3.25--4.75) & 3.83 (1.17)
& 4.00 (4.00--4.00) & \textbf{4.20 (0.42)} \\
Q1.3 & The integrated LLM helped me reach a satisfactory conclusion.
& 4.00 (4.00--4.00) & 3.83 (0.98)
& 4.00 (4.00--4.00) & \textbf{4.10 (0.57)} \\
Q1.4 & The sketching interface helped me think clearly and easily.
& 4.50 (2.50--5.00) & 3.83 (1.47)
& 5.00 (5.00--5.00) & \textbf{4.80 (0.42)} \\
Q1.5 & Using this interface, I was able to make better progress toward my goal.
& --- & ---
& 5.00 (4.00--5.00) & \textbf{4.50 (0.71)} \\

\rowcolor{black}
\multicolumn{2}{l}{\textcolor{white}{\textbf{User Interface and Interaction Design}}} 
& & \textcolor{white}{3.54} 
& & \textcolor{white}{\textbf{3.90}} \\
\midrule
Q2.1 & It was easy to convey my intent to the AI using sketches and handwriting.
& 4.00 (3.25--4.00) & 3.67 (1.03)
& 4.00 (4.00--4.75) & \textbf{3.90 (1.10)} \\
Q2.2 & The AI can successfully interpret my intention in the spatial arrangement of my drawings.
& 4.00 (3.25--4.00) & \textbf{3.83 (0.75)}
& 3.50 (3.00--4.00) & 3.70 (0.82) \\
Q2.3 & I did not feel like I had to fight the tool to make the AI understand me.
& 4.00 (2.50--4.75) & 3.67 (1.37)
& 4.00 (4.00--4.00) & \textbf{3.90 (0.74)} \\
Q2.4 & The inking experience mimics the feel of pen on paper.
& 3.00 (2.00--4.00) & 3.00 (1.55)
& 4.00 (2.25--4.75) & \textbf{3.50 (1.43)} \\
Q2.5 & It was clear and easy to navigate between different screens of the app, including note-taking and AI modes.
& --- & ---
& 4.00 (4.00--5.00) & \textbf{4.40 (0.52)} \\
Q2.6 & The interface provided enough visible cues (labels, icons, or feedback) for me to confidently find and use features.
& --- & ---
& 5.00 (4.00--5.00) & \textbf{4.30 (1.06)} \\
Q2.7 & I had enough control to adjust, undo, or override the system's merging/grouping of my notes.
& --- & ---
& 4.00 (3.25--4.00) & \textbf{3.60 (1.26)} \\

\rowcolor{black}
\multicolumn{2}{l}{\textcolor{white}{\textbf{AI Integration}}} 
& & \textcolor{white}{---} 
& & \textcolor{white}{\textbf{4.40}} \\
\midrule
Q3.1 & AI responses appeared at appropriate times and did not interrupt or distract from my thinking process.
& --- & ---
& 4.50 (4.00--5.00) & \textbf{4.40 (0.70)} \\

\rowcolor{black}
\multicolumn{2}{l}{\textcolor{white}{\textbf{AI Outputs}}} 
& & \textcolor{white}{3.83} 
& & \textcolor{white}{\textbf{4.12}} \\
\midrule
Q4.1 & The AI's responses were useful to the context of my problem.
& 4.00 (4.00--4.00) & 4.17 (0.41)
& 4.00 (4.00--4.75) & \textbf{4.30 (0.48)} \\
Q4.2 & The presentation/format of the AI's response as text on the canvas was useful to the context of my problem.
& 4.50 (2.50--5.00) & 3.83 (1.47)
& 5.00 (4.25--5.00) & \textbf{4.50 (0.85)} \\
Q4.3 & The AI's responses appeared in the location that I expected them to.
& 4.00 (2.25--5.00) & 3.50 (1.76)
& 5.00 (4.00--5.00) & \textbf{4.20 (1.23)} \\
Q4.4 & I could understand and control what information the AI used from my canvas when generating a response.
& --- & ---
& 3.50 (3.00--5.00) & \textbf{3.70 (1.25)} \\
Q4.5 & It was easy to revisit, recover, and compare previous AI generations when I needed them.
& --- & ---
& 4.00 (4.00--4.00) & \textbf{3.90 (0.88)} \\

\bottomrule
\end{tabularx}
\vspace{1mm}
\caption{Combined descriptive statistics for post-task questionnaires. Categorical headers display aggregate mean scores. Items were rated on a 5-point Likert scale. Bold indicates the higher Mean (SD) between studies for that row; where only one study was available, its Mean (SD) is bolded.}
\end{table*}

%TC:endignore 

\end{document}